\documentclass[aps,prb,reprint,superscriptaddress,showpacs]{revtex4-1}

\usepackage[english]{babel}
\usepackage{ragged2e}
\justifying
\usepackage{amsmath}
\usepackage{tikz}
\usetikzlibrary{shapes,arrows}
\usepackage{verbatim}
\usepackage{graphicx}
\usepackage{float}
\usepackage[caption=false]{subfig}
\usepackage{amsfonts}
\usepackage{amssymb}
\usepackage{hyperref}

\begin{document}

\title{Classical and quantum dynamics of indirect excitons driven by surface acoustic waves}

\author{Federico Grasselli}
\email[]{federico.grasselli@unimore.it}
\affiliation{International School for Advanced Studies, Via Bonomea 265, Trieste, Italy}

\author{Andrea Bertoni}
\email[]{andrea.bertoni@nano.cnr.it}
\affiliation{S3, Consiglio Nazionale delle Ricerche, Istituto Nanoscienze, Via Campi 213/a, Modena, Italy}

\author{Guido Goldoni}
\email[]{guido.goldoni@unimore.it}
\affiliation{Dipartimento di Scienze Fisiche, Informatiche e Matematiche, Universit\`a degli Studi di Modena e Reggio Emilia, Via Campi 213/a, Modena, Italy}
\affiliation{S3, Consiglio Nazionale delle Ricerche, Istituto Nanoscienze, Via Campi 213/a, Modena, Italy}

\date{\today}

\begin{abstract}
We perform explicit time-dependent classical and quantum propagation of a spatially indirect exciton (SIX) driven by surface acoustic waves (SAWs) in a semiconductor heterostructure device. 
We model the SIX dynamics at different levels of description, from the Euler-Lagrange propagation of structureless classical particles to unitary Schr\"odinger propagation of an electron-hole wave packet in a mean field and to the full quantum propagation of the two-particle complex. 
A recently proposed beyond mean-field self-energy approach, adding internal virtual transitions to the c.m. dynamics, has been generalized to time-dependent potentials and turns out to describe very well full quantum calculations, while being orders of magnitude numerically less demanding.
We show that SAW-driven SIXs are a sensitive probe of scattering potentials in the devices originating, for example, from single impurities or metallic gates, due to competing length and energy scales between the SAW elastic potential, the scattering potential, and the internal electron-hole dynamic of the SIX. 
Comparison between different approximations allow to show that internal correlation of the electron-hole pair is crucial in scattering from shallow impurities, where tunneling plays a major role. On the other hand, scattering from broad potentials, i.e. with length scales exceeding the SIX Bohr radius, is well described as the classical dynamics of a point-like SIX. Recent experiments are discussed in light of our calculations. 
\end{abstract}


\maketitle

\section{Introduction}

Surface acoustic waves (SAWs) in semiconductors, with typical wavelength $\sim 1\mu\mbox{m}$, can be controllably generated by interdigital devices.\cite{Datta_SAW86,Kovalev_JETPL15,Weiss_PRApp18}
SAWs travel for microns without attenuation and couple with charge carriers through their strain field.
Recently, the SAW field has been shown to trap and transport spatially indirect excitons\cite{ViolanteNJP14} (SIXs), photoexcited electron-hole pairs with the two charges held in different layers of a coupled-quantum-well (CQW) structure by a vertical electric field\cite{High_Science11072008,Sivalertporn_PRB12,AndreakouPRB15,Kovalev_JETP16,BeianPRAPP17}. 
With respect to single-layer excitons, the spatial separation of the two charges by the interlayer separation $d$ (a few nm) donates to SIXs a much extended lifetime (in the $\mu$s range) and a finite dipole moment $ed$\cite{WilkesPRB16,Kuznetsova_PRB17} which allows SIXs to be accelerated by a potential gradient along the planes of the wells. 
This may be a static potential, as generated by an extended impurity or a metallic gate grown on top of the CQW device, or a time-dependent (TD) potential, as generated by the SAWs strain field.

Differently from piezoelectric SAWs\cite{AstleyPRL07,KataokaPRL09,BuscemiJPCM09,McNeil_Nature11}, where exciton lifetime is due to the electron-hole \emph{in-plane} separation \emph{along the plane} induced by the type-II SAW field itself, SIXs are  intrinsically long-lived due to the \emph{vertical} separation. Therefore, SIXs are well defined quasi-particles inside non-piezoelectric\cite{Rudolph_PRL07,Rudolph_ProcCPLMCN07}, type-I SAWs. Moreover, SIX space- and time-propagation can be optically probed, since recombination in the nanosecond timescale can be induced 'on demand' by switching off the vertical electric field. As a result, SAW-driven SIXs (SAW-SIXs) may be i) used to store information, say, from a laser pulse, ii) controllably driven across the device, possibly performing a quantum gate operation,\cite{RosiniJCE2004,BordoneSST04,RodriquezPRB2005,BuscemiPRB10,ShiPRA11} and iii) read off the result of the computation inducing optical recombination by compensating the vertical field. This class of \textit{excitonic devices} couples the advantage of long distance communication with light and fast electronic switching.\cite{Grosso_NatPhot09} 

In a different perspective, SAW-SIXs traveling for large distances with fixed group velocity $v_G$, can be considered as particles which, being extended over about an effective Bohr radius, interact with impurities or other potential modulations of comparable length scale.
Accordingly, SAW-SIXs may probe short-range potential landscapes through a scattering event.
Probing local electrostatic potential in semiconductors is a primary goal, e.g., to characterize type and position of single impurities, extended or lattice defects, interface modulations, etc. 
This is often indirectly performed by exciton binding, e.g., to impurities, to give spectroscopic fingerprints of the impurity itself and its environment. 
Type-I SAWs have been recently used to extract disorder-limited mobilities in nanowires\cite{KinzelACSNano16}, but the high sensitivity of SIXs to local imperfections of the hosting material, leading to the coupling of c.m. and internal degrees of freedom (DOFs) and eventually to SIXs recombination, may become an additional way to probe a target local potential.

Most theoretical investigations of SIXs dynamics treat these quasi-particles as rigid, single-particle objects, neglecting the internal DOFs. This is a severe simplification in several instances, when the external potentials induce energy transfer between c.m. and relative DOFs during scattering events.\cite{GrasselliJCP15,GrasselliPRB16,GrasselliJPCS16}
Even in cases where the energy scale of the external potential is too small with respect to the internal gaps of the SIXs (a few meV) to induce internal excitations, virtual transitions may induce a phase shift which strongly renormalizes transmission resonances or tunneling probabilities. 
This has been shown, in particular, by a self-energy approach recently developed by the authors which, restoring the effects of virtual transitions in the c.m.~dynamics by an energy-dependent potential, gives very good agreement with exact two-particle Schr\"odinger propagation which takes fully into account internal electron-hole correlations.\cite{GrasselliPRB16b, GrasselliSLMS17}

In this paper we investigate the dynamics of SAW-SIXs scattering against selected potential landscapes in different experimentally-relevant regimes. We perform exact quantum evolution in space and time of SAW-SIXs wave packets, taking fully into account internal DOFs. 
Comparing exact results with mean-field and our self-energy approach,\cite{GrasselliPRB16b} which is here extended to time-dependent potentials, we single out the role of internal correlations. As show for static potentials,\cite{GrasselliPRB16b} within the self-energy approach correlations correctly accounted for with orders-of-magnitude reduced computational load with respect to full quantum propagation. Finally, Euler-Lagrange propagation of structureless classical particles exposes the dominant role of quantum tunneling effects in scattering against sharp and localized external potentials. Recent experiments are discussed in light of our calculations. 

In Sec.~\ref{sec:PhysicalSystem} we describe the theoretical framework of the problem, defining the different models to describe SAW-SIX propagation, at different levels of approximation.
In Sec. \ref{sec:Results} we report the main results obtained through specific simulations, performed in paradigmatic and experimentally relevant regimes. 
In Sec. \ref{sec:Conclusions} we draw the main conclusions of this work.
In the Appendices, we give further details on the representation in terms of c.m. and relative coordinates are given (App. \ref{App:Xxrepresentation}), and on the periodic motion of a classical SAW-SIX (App. \ref{App:ClassPeriodMot}).

\section{Theoretical description}\label{sec:PhysicalSystem}

\subsection{Physical system}

We simulate SIXs transport over several $\mu$m induced by a \textit{non-piezoelectric} SAW which propagates parallel to the planes of a CQW heterotructure. A non-piezoelectric SAW generates type-I strain-induced modulation of the valence and the conduction bands (VB and CB, respectively), i.e., the confinement of the electron and of the hole is induced at the same in-plane position, corresponding to the minimum of the CB and maximum of the VB.
This strongly stabilizes the relative motion of the pair, preventing the exciton from dissociation in the absence of further external potentials. A scheme of the double-well system, with the SIX bound in the minima of the type-I SAW is reported in the inset of Fig.~\ref{fig:SAW:figure1} 
The distance of the CQW layer from the top of the heterostructure can be engineered so that the type-I potential profile induces almost equal modulations in the VB and CB amplitudes in the meV range. 
Below, we consider a SAW with a wavelength $\lambda_\mathrm{SAW} = 2.8\,\mathrm{\mu m}$ and a period $T \sim 1\,\mathrm{ns}$ corresponding to a frequency of $\approx 1\,\mathrm{GHz}$ at the temperature of 4 K (see Ref.~\onlinecite{ViolanteNJP14}).

We shall analyze the dynamics of SAW-SIXs scattering against selected (Gaussian) external potentials, mimicking different types of scattering potentials with different strength and using several propagation methods. In particular, we shall compare quantum propagation of wave packets at different levels of approximation and classical propagation of point-like excitons.

\subsection{Quantum simulation}
   
\subsubsection{Hamiltonian and initial state}

We consider a 1D electron-hole pair, with coordinates $x_e, x_h$ and effective masses $m_e, m_h$, for the electron and the hole, respectively, each particle being described in a single-band model.
The SIX complex has c.m.~and relative coordinates  $X = (m_e x_e + m_h x_h )/M$ and $x \equiv x_e - x_h $, and total and reduced masses $M = m_e+m_h$ and $m = (m_e^{-1} + m_h^{-1})^{-1}$, respectively. 
In the absence of any external potential the c.m.~and internal DOF separate, and the internal SIX eigenstates $\phi_n$ are determined by the relative electron-hole Hamiltonian 
\begin{equation}
H_\mathrm{rel} = \frac{p^2}{2m} + U_{\mathrm{int}}(x),\label{eq:freerelHam}
\end{equation}
where $U_{\mathrm{int}}(x)$ is the electron-hole interaction, and $p$ the relative momentum. 
For the calculations we take
\begin{equation}
U_{\mathrm{int}}(x) = -\frac{e^2}{4\pi\epsilon_0\epsilon_r} \frac{1}{\sqrt{x^2 + d^2}},
\end{equation}
where $\epsilon_0$ is the dielectric permittivity in vacuum, $\epsilon_r$ is the relative dielectric permittivity of the material, and $d$ is the vertical separation between the two charges, assumed to be the center-to-center separation of the two quantum wells.
Unless differently stated, we take $\epsilon_r =12.9$ and $d=20\,\mathrm{nm}$, as these are typical parameters of many GaAs-based CQW devices,\cite{GrasselliPRB16b, GrasselliSLMS17} which are used, e.g., for SAW-SIXs.\cite{ViolanteNJP14}
The external potential 
\begin{equation}
U_{\mathrm{ext}} = U_{e}(x_e;t) + U_{h}(x_h;t)
\end{equation}
includes both the time-dependent SAW and the stationary impurity or gate-generated potential,
acting on each particle separately. 
Therefore, in relative and c.m. coordinates, the total Hamiltonian reads
\begin{equation}
H_{\mathrm{tot}} = \frac{P^2}{2M} +H_\mathrm{rel} + U_{\mathrm{ext}}(X,x;t) \,,\label{eq:genHam}
\end{equation}
where $P$ is the c.m. momentum.
Since $U_{\mathrm{ext}}$ removes translational invariance, c.m.~and internal DOFs are coupled. $U_{\mathrm{ext}}$ includes \emph{both} the SAW field $U_\mathrm{SAW}$ and any scattering potential $U_\mathrm{scat}$,
\begin{equation}
U_{\mathrm{ext}} = U_\mathrm{SAW} + U_\mathrm{scat}\,.
\end{equation}

Note that $U_e$ and $U_h$ are usually different. 
First, the external field generated, e.g., by a charged impurity or a surface metallic gate results in a potential energy which has opposite \emph{sign} for the two particles. 
Second, since electrons and holes are confined to different layers, also the \emph{intensity} of the generated potential may be different, this difference being possibly in the order of the internal excitations.
Therefore, in general the two contributions do not cancel out. Therefore, below we model the scattering potentials by a Gaussian shape, possibly with different parameters for the electron and the hole component. 
This may account qualitatively for several classes of external potentials, such as created by electrostatic gates, impurities, etc.~by appropriate choices of the parameters, 

In practice, for numerical convenience, we study the dynamics of SIXs into the moving reference frame with the SAW at rest. Therefore, we model the SAW potential 
\begin{equation}
U_{\mathrm{SAW}}(X,x) = U_{\mathrm{SAW},e} + U_{\mathrm{SAW},h}
\end{equation} 
as a sinusoidal wave 
\begin{equation}
U_{\mathrm{SAW},e(h)} = U_{e(h),0} \sin(2\pi x_{e(h)}/\lambda_\mathrm{SAW})
\end{equation}
for the electron (the hole). Here $\lambda_\mathrm{SAW}$ is the SAW wavelength, and $U_{e(h),0}$ the amplitude of the periodic modulation of the CB (VB), at the electron (hole) QW location. 
In the SAW reference frame, any stationary external potential moves with velocity $-v_\mathrm{SAW}$. Therefore,
\begin{equation}
U_\mathrm{scat} = U^\mathrm{G}_e(x_{e};t) + U^\mathrm{G}_h(x_{h};t) \, ,
\end{equation}
where
\begin{equation}
U_{\alpha}^\mathrm{G}(x_{\alpha};t) = U_{\alpha,0}^\mathrm{G} \exp\left[-\frac{(x_{\alpha} - x_0^\mathrm{G} -v_\mathrm{SAW}t)^2}{2\sigma_\mathrm{G}^2}\right] \, .
\label{eq:GaussPot}
\end{equation}
Here $\alpha=e,h$, $x_0^\mathrm{G}$ is the initial center of the Gaussian, $\sigma_\mathrm{G}$ its localization length, $|U_{\alpha,0}^\mathrm{G}|$ its amplitude.
For simplicity we assume the same $\sigma_\mathrm{G}$ and $x_0^\mathrm{G}$, but different intensities, for the two particles.

\begin{figure}
\begin{center}
\includegraphics[width=0.9\columnwidth]{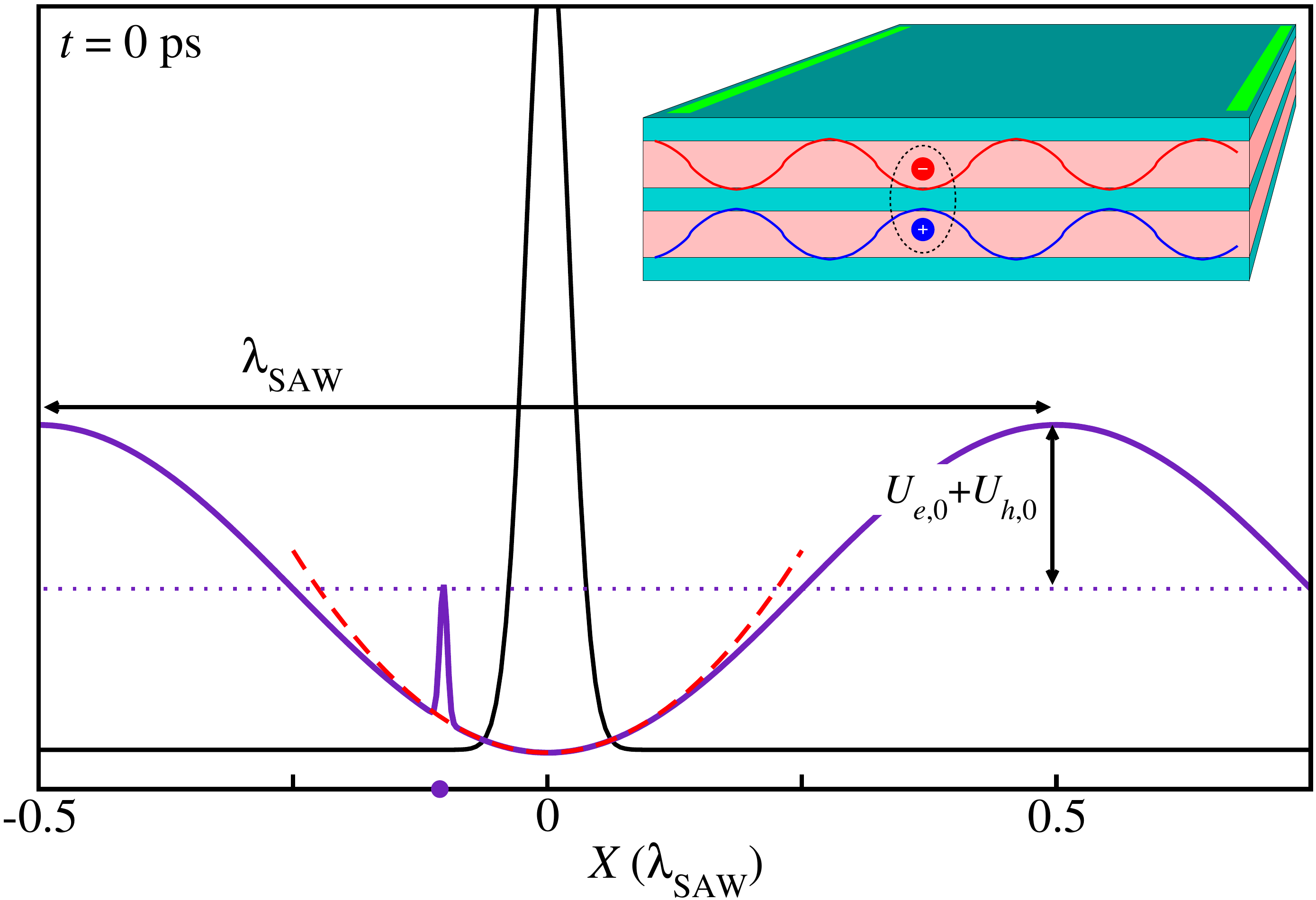}
\end{center}
\caption{\small Initial c.m. probability distribution  $|\chi_\mathrm{HO}(X)|^2$ (black solid line), and external potential (violet solid line), sum of the SAW and a Gaussian scattering potential, taken at $x=0$. The parabolic approximation at a SAW minimum is displayed (red dashed line). 
Inset: schematic illustration of the double quantum well device where the SIX is bound in the minimum of a type-I SAW.\label{fig:SAW:figure1}}
\end{figure}

To initialize the quantum simulation, we notice that the lateral modulation of the SAW is shallow on the length scale of the SIX extension: $\lambda_\mathrm{SAW}\sim\mu m$, while the SIX effective Bohr radius is $a_\mathrm{B}^* \sim 10\,\mathrm{nm}$. 
Therefore, we assume that SIXs relax to the ground state of the harmonic potential which approximates a minimum of the SAW, and we choose \begin{equation}
\Psi(X,x,t=0) = \chi_\mathrm{HO}(X)\phi_0(x) \, ,
\label{eq:InitialState}
\end{equation}
where $\phi_0(x)$ is the ground state of the free relative Hamiltonian, $H_\mathrm{rel}$ in Eq.~(\ref{eq:freerelHam}), and $\chi_\mathrm{HO}(X)$ is the c.m.~ground state of the harmonic approximation of the SAW potential about one of its minima, denoted by $X_0$, and at $x\approx 0$
\begin{equation}
\left.U_\mathrm{SAW} (X)\right|_{X\approx X_0, x \approx 0} \approx  |U_{e,0} + U_{h,0}|\left(-1 + \frac{k_\mathrm{SAW}^2}{2}X^2 \right).
\end{equation}
Therefore,
\begin{equation}
\chi_\mathrm{HO}(X) = \left(\frac{M\omega_\mathrm{HO}}{\hbar \pi}\right)^{1/4} \exp\left[-\frac{(X-X_0)^2}{4\sigma_\mathrm{HO}^2}\right] \, ,
\end{equation}
where $\omega_\mathrm{HO}\equiv k_\mathrm{SAW} \sqrt{|U_{e,0} + U_{h,0}|/M}$ and $\sigma_\mathrm{HO}^2\equiv \hbar/({2k_\mathrm{SAW}\sqrt{M|U_{e,0} + U_{h,0}|}})$.
Figure~\ref{fig:SAW:figure1} shows the square modulus of the c.m.~component of the ground state, $|\chi_\mathrm{HO}(X)|^2$ (black solid line), as well as the potential profile, at $x=0$, of the sum of the SAW potential and the Gaussian scattering potential (violet solid line). A violet spot highlights the position of the center of the Gaussian potential. The parabolic approximation nearby the SAW minimum is also displayed (red dashed line). 

\subsubsection{Exact quantum propagation}
To perform exact TD propagation of an electron-hole wave packet $\Psi(X,x,t)$ we solve explicitly the two-particle equation of motion 
\begin{equation}
i\partial_t \Psi(X,x,t) = H_\mathrm{tot} \Psi(X,x,t) .
\end{equation}
We start from the initial state $\Psi(X,x,t=0)$, given by Eq.~(\ref{eq:InitialState}) and simulate the repeated application of the unitary evolution operator $ \mathcal{U}(t+\Delta t,t)$ between times $t$ and $t + \Delta_t$, 
\begin{equation}
\Psi(X,x;t+\Delta_t) = \mathcal{U}(t+\Delta_t,t) \Psi(X,x;t)\,.
\label{eq:propagation}
\end{equation} 
Using the Fourier split-step (FSS) method,\cite{GrasselliJCP15,GrasselliPRB16} 
\begin{equation}
\mathcal{U}(t+\Delta_t,t) = e^{-\frac{i}{\hbar}U(t+\frac{\Delta_t}{2})\frac{\Delta_t}{2}}e^{-\frac{i}{\hbar}T\Delta_t}e^{-\frac{i}{\hbar}U(t+\frac{\Delta_t}{2})\frac{\Delta_t}{2}} \, ,
\end{equation}
which is second-order accurate in $\Delta_t$.
Here $U=U_\mathrm{int}+U_\mathrm{ext}$ and $T=P^2/(2M) + p^2/(2m)$ are the total potential- and kinetic-energy operators.

Since $\mathcal{U}(t+\Delta_t,t) $ represents in full the two-particle dynamics, including both external and internal interactions of the electron-hole pair, it provides a (numerically) \emph{exact} unitary propagation of a SAW-SIX. We stress that, even though the initial wave packet is factorized, no assumption is  made \textit{a priori} on the wave-packet form during the propagation. Therefore, all dynamical correlations of the pair during scattering with the external potential are taken into account.

Later we will analyze the results of the TD evolution in terms of the marginal probability 
\begin{equation}
{\rho_\mathrm{cm}}(X;t)\equiv \int dx |\Psi(X,x;t)|^2,
\end{equation}
which can be understood as the c.m.~probability density.

\subsubsection{Point-like exciton}
At the lowest level, in the scale of approximate descriptions of quantum SIXs, there is a point-like model where the internal structure of the exciton is totally disregarded in the quantum propagation\cite{Lobanov_PRB16}. This is equivalent to taking the internal motion wave function as a Dirac delta centered at $x=0$, $\Psi(X,x,t)=\chi(X,t)\delta(x)$.
The c.m.~wave function $\chi(X,t)$ evolves according to 
\begin{equation}
i\hbar\partial_t \chi(X) = H_\mathrm{cm} \chi(X,t)\,.
\label{eq:pointlike}
\end{equation}
with 
\begin{equation}
H_\mathrm{cm} = \frac{P^2}{2M}+ U_\mathrm{cm}(X,t)\,.
\end{equation}
and the effective potential  $U_\mathrm{cm}(X,t) = U_\mathrm{ext}(X,x=0;t)$.

As in the exact propagation, we use the FSS method with 
\begin{equation}
\mathcal{U}(t+\Delta_t,t) = e^{-\frac{i}{\hbar}U_\mathrm{cm}(t+\frac{\Delta_t}{2})\frac{\Delta_t}{2}}e^{-\frac{i}{\hbar}\frac{P^2}{2M}\Delta_t}e^{-\frac{i}{\hbar}U_\mathrm{cm}(t+\frac{\Delta_t}{2})\frac{\Delta_t}{2}}
\end{equation}
acting on the c.m.~wave function $\chi(X;t)$ at each time step.

\subsubsection{Rigid exciton approximation}

In semiconductor systems, if external potentials are sufficiently smooth on the scale of the effective Bohr radius (here about 10 nm), an accurate representation of the excitonic states is to take the SIX envelope function factorized as
\begin{equation}
\Psi(X,x;t) = \chi(X;t)\phi_0(x) \, ,
\end{equation}
where $\phi_0(x)$ is the ground state of the relative motion Hamiltonian, Eq.~(\ref{eq:freerelHam}). 
When substituted in the full Hamiltonian, the two-body equation reduces to Eq.~(\ref{eq:pointlike}) but with effective external potential 
\begin{equation}
U_\mathrm{cm}(X;t) \equiv U_\mathrm{RIX}(X;t) = \int U_\mathrm{ext}(X,x;t) |\phi_0(x)|^2 dx\,,
\end{equation} 
i.e., the expectation value of the total external potential onto the internal (free) ground state, $\phi_0$.  

This is called the \textit{rigid exciton} (RIX) approximation (see Refs.~\onlinecite{ZimmermannPAC97, GrasselliPRB16}), as it assumes that the electron-hole pair is frozen in the relative motion ground state during the propagation.  

\subsubsection{Self-energy approach}\label{sec:SEapproach}

Recently, we improved the RIX approximation by restoring virtual transitions to higher states $\phi_m(x)$ of Eq.~(\ref{eq:freerelHam}), with eigenenergy $\epsilon_m$, through a properly designed, local and energy-dependent self--energy, $\Sigma$.\cite{GrasselliPRB16b} 
For low frequency TD external potentials, this reads 
\begin{equation}
\Sigma(X;E(t);t) = \sum^{N_c}_{m\neq 0} \frac{|W_{0 m}(X;t)|^2}{E(t)-\epsilon_m}\,,
\label{eq:SE1Dfree} 
\end{equation}
where
\begin{equation}
W_{0 m}(R;t) \equiv \int \phi_0(x)^\ast U_\mathrm{ext}(X,x;t) \phi_m(x)dx \, , 
\end{equation}  
$E(t)$ is the total energy of the system, which for TD external potentials is time dependent, and $N_c$ is a cutoff number of internal states, sufficient to reach proper convergence on $\Sigma$.

Within this approximation, the c.m.~wave function propagates under the effective potential, 
given by
\begin{equation}
 U_\mathrm{cm}(X;t) = U_\mathrm{RIX}(X;t) + \Sigma(X;E(t);t)\,. 
\end{equation} 
Note that, similarly to the RIX approximation, only the c.m. coordinates need to be propagated in a potential landscape generated by the external scattering potential, but renormalized by the electron-hole interaction. Note also that $ \Sigma(X;E(t);t)$ can be calculated from the knowledge of the \emph{free} -- i.e., in the absence of any scattering potential--SIX.

\subsection{Classical simulation}

Often in literature the quantum-mechanical nature of the SIXs is neglected, like in drift and diffusion models, where the quantum (bosonic) nature of IXs enters into the simulations only at the statistical level through the Bose-Einstein distribution function. \cite{Combescot_RepProgPhys17} 
In order to single out the role of the quantum nature of SIXs we also solved the Euler-Lagrange equations of motion for structureless \textit{classical} IXs, moving in the potential landscape generated by the SAW and the scattering  potentials. Specifically, assuming Gaussian scattering potentials as in Eq.~(\ref{eq:GaussPot}), the Lagrangian reads
\begin{equation}
\begin{split}
L(X,\dot{X};t) &= \, \frac{1}{2}M \dot{X}^2 \\
&- U_0^\mathrm{SAW}\sin[k_\mathrm{SAW} (X-X_0)] \\
&- U_0^\mathrm{G} \exp \left[-\frac{(X - x_0^\mathrm{G} -v_\mathrm{G}t)^2}{2\sigma_\mathrm{G}^2}\right],
\end{split}
\end{equation}
where $U_0^\mathrm{SAW} \equiv U_{e,0}^\mathrm{SAW} + U_{h,0}^\mathrm{SAW}$, and $U_0^\mathrm{G} \equiv U_{e,0}^\mathrm{G} + U_{h,0}^\mathrm{G}$.

An issue arises in the choice of the initial conditions, since the initial c.m.~position and velocity of the IX are not well determined for the initial wave packet chosen as initial state, $\chi_\mathrm{GS}(X)$, and a correspondence between the initial quantum probability density and the initial classical conditions must be established. In order to compare classical and quantum simulations, we solve the equations of motion for a \textit{distribution} of initial conditions, $[X_i(0),\dot{X}_i(0)]$, with $i=1,\ldots,N$, where $N$ is the total number of classical trajectories. $X_i(0)$ and $\dot{X}_i(0)$ are randomly chosen according to the distributions provided by $|\chi_\mathrm{GS}(X)|^2$, and its Fourier transform, $|\tilde{\chi}_\mathrm{GS}(P/M)|^2$, respectively.

\section{Numerical Results}\label{sec:Results}

\subsection{Scattering by a shallow impurity}

We consider a narrow Gaussian scattering potential with $\sigma_\mathrm{G} = 10 \, \mathrm{nm} \sim a_\mathrm{B}^*\ll \lambda_\mathrm{SAW}$ to mimic a shallow ionized impurity. We shall perform simulations in different regimes with the intensity of the impurity potential from smaller to comparable to the SAW amplitude. 

The wave packet is initialized in a minimum of the SAW potential, at the right hand side of the scattering potential, as shown in Fig.~\ref{fig:SAW:figure1} together with the total potential (violet), composed of the SAW and scattering potentials. 
We remark that in our coordinate system, the SAW potential is at rest, while the scattering potential moves to the right. 

\begin{figure*}
\includegraphics[width=\textwidth]{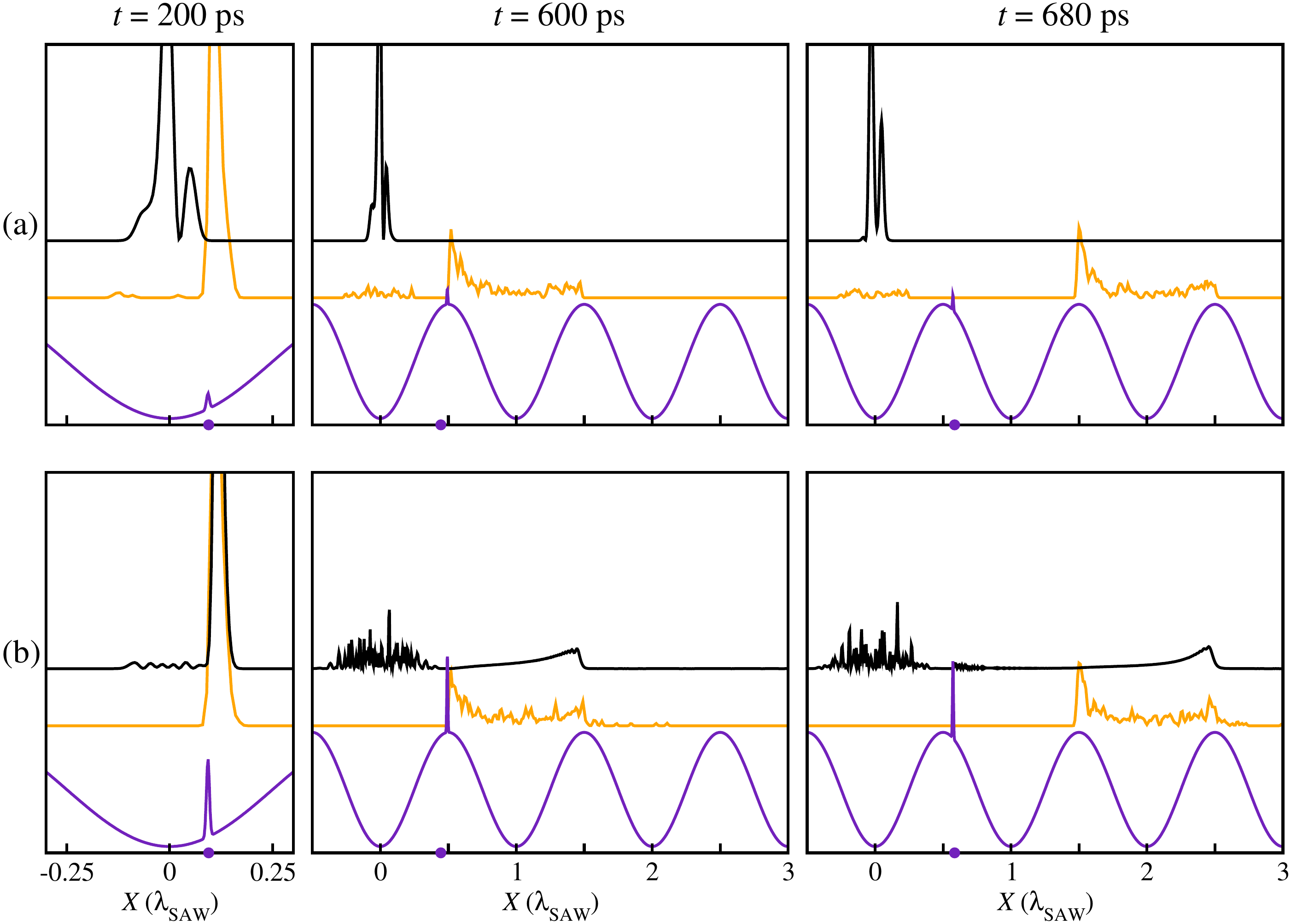}
\caption{Propagation in the presence of an external potential with Gaussian shape and with two different maximum values: (a) $U_{e,0}^\mathrm{G}=+1.5$~meV; (b) $U_{e,0}^\mathrm{G}=+3.5$~meV.  Fixed are $U_{h,0}^\mathrm{G}=-1.0$~meV and $\sigma_G=10$~nm.  Displayed curves: exact calculation (black), classical distribution (orange). $U_\mathrm{ext}(X,x=0;t)$ is shown as a violet line. Snapshots at three different times. The initial state is described in Fig.~\ref{fig:SAW:figure1}. A violet dot on the $X$ axis indicates the center of the fixed external potential. Here and in the following, wave packets computed with different methods are shifted on the $Y$ axis by a fixed offset for clarity.
\label{fig:SAW:figure2}}
\end{figure*}

In Fig.~\ref{fig:SAW:figure2} we show the time evolution (time increases from left to right) of the c.m.~marginal probability, $\rho_\mathrm{cm}(X;t)$, obtained via exact (black line), and classical (orange line) propagation. In the two panels of Fig.~\ref{fig:SAW:figure2}, the SAW sinusoidal potentials have intensities $U_{e,0}=U_{h,0}=0.9$~meV. A Gaussian external scattering potential as given by Eq.~(\ref{eq:GaussPot}), is added, with $U_{h,0}^\mathrm{G}=-1.0$~meV, and increasing values of $U_{e,0}^\mathrm{G}=+1.5$ and $3.5$~meV in Figs.~ \ref{fig:SAW:figure2}(a) and (b), respectively. 
Figure \ref{fig:SAW:figure4} reports several calculation methods for an intermediate regime, with $U_{e,0}^\mathrm{G}=+2.5$,
Again, the total external potential is shown as a violet line in each panel, the central position of the Gaussian external potential being highlighted by a violet dot. 

Since in our simulations we neglect relaxation processes (e.g. mediated by phonons), high Fourier components are visible and persist at times longer than typical relaxation times. Although the form of the wave function would be different when relaxation is considered, its localization in the different SAW minima is not expected to be substantially affected.

\subsubsection{Weak scattering potentials}

For a weak impurity potential, $U_{e,0}^\mathrm{G}=+1.5$~meV [Fig.~\ref{fig:SAW:figure2}(a)], most of $\rho_\mathrm{cm}(X;t)$ remains localized into the initial SAW minimum, since the SIX tunnels through the Gaussian potential. 
Note that at $T=200\,\mbox{ps}$ the tunneled wave packet is composed of a few Fourier components. The wave packet distribution is the result of the interference of one reflection inside the SAW.
At later times, multiple reflections rebuild a state inside the initial SAW minimum, similar to the initial state but characterized by more peaks, and on the opposite side of the scattering potential.

Since this is a purely quantum effect, it is at striking difference with the classical simulation, where only a small fraction of the classical distribution remains localized into the initial SAW potential well. Only few classical trajectories, with energy larger than $U_{h,0}^\mathrm{G} + U_{e,0}^\mathrm{G} = 0.5$~meV, may overcome the external potential. All other trajectories are squeezed against the scattering potential by the SAW until the energy is large enough to surmount the scattering barrier. Other particles are released to the right when the SAW maximum is at the same position of the scattering potential, having accumulated from the SAW a potential energy up to the double of the amplitude of the SAW potential, with an average velocity $\lambda_\mathrm{SAW}/T \gg V_\mathrm{SAW}$, where $T$ is the period to travel from one crest to the following one -- see Sec.\ref{App:ClassPeriodMot}.
Essentially, the SIX undergoes an adiabatic inelastic scattering and accumulates enough kinetic energy to propagate at a mean velocity larger than the SAW velocity. As stated before, this effect is quenched in the quantum case due to the possibility of the SIX to tunnel through the impurity potential.
This quantum effect should be measurable by means of the luminescence pattern, reflecting the spatial localization of the SIX: while classical particles acquires a kinetic energy of the order of the SAW amplitude and do not remain in the original SAW minimum, quantum particles, able to tunnel the impurity potential, stay in the original SAW. The rightmost panel of Fig.~\ref{fig:SAW:figure2}(a) illustrates the difference.

\subsubsection{Strong scattering potential}\label{sec:strongpots}

In Fig.\ref{fig:SAW:figure2}(b), the scattering intensity $U_{e,0}^\mathrm{G}$ is twice the SAW amplitude. 
In this case, the tunneling process takes a longer time, but complete tunneling occurs when the SAW has totally crossed the scattering potential (see panel at $t=600s$). 
Again, this differs from the classical trajectory distribution, which is similar to the previous case except that here, due to the stronger potential, no classical trajectory gains sufficient energy to overcome the barrier and no trajectory is found in the original SAW minimum on the left hand side ($X=0$), the whole distribution being reflected to the right. 

\begin{figure*}
\includegraphics[width=\textwidth]{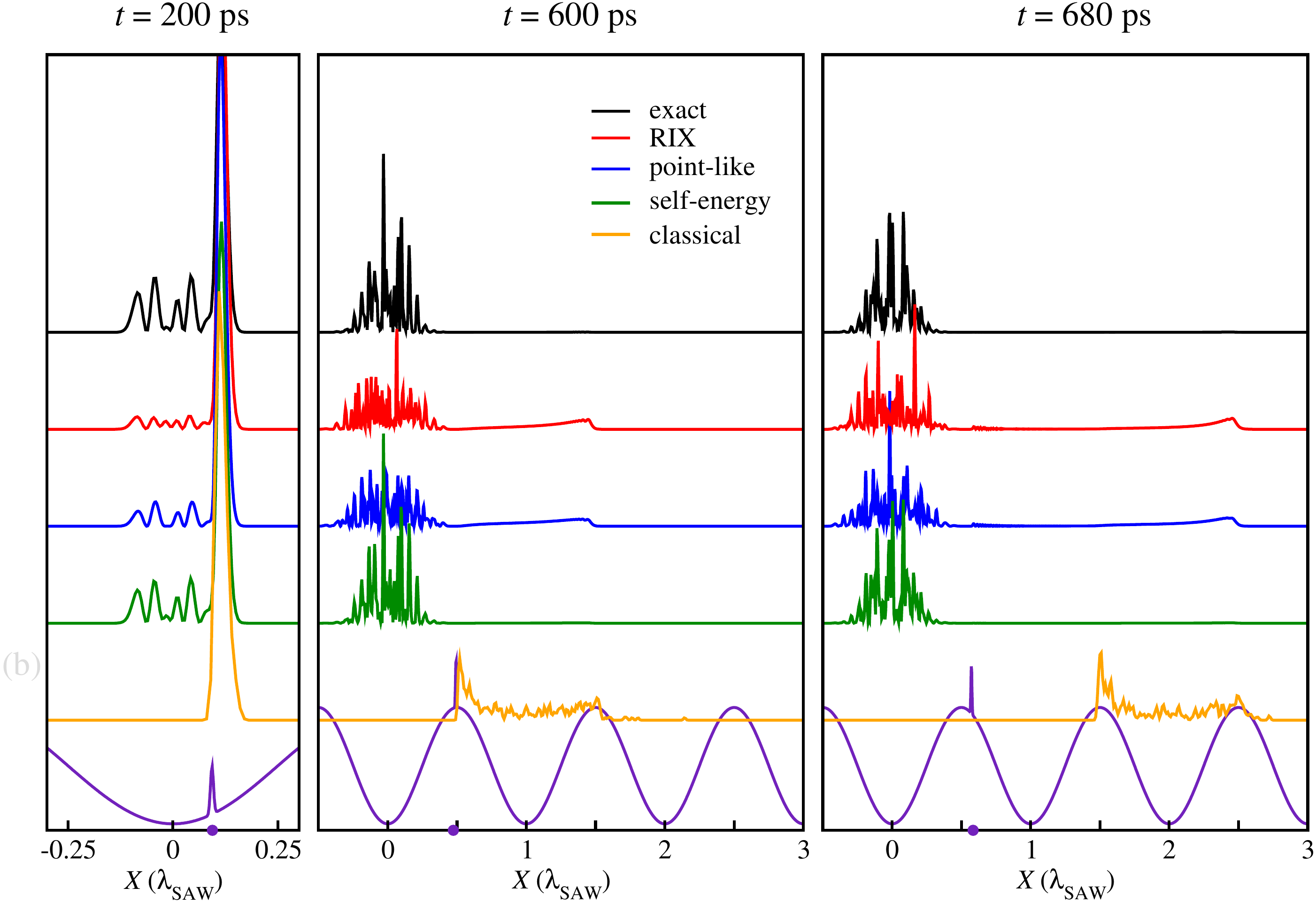}
\caption{Comparison of exact (black), rigid-exciton (red), point-like particle (blue), self-energy (green), and classical-particle (orange) methods, for an intermediate regime with respect to the propagations presented in Fig.~\ref{fig:SAW:figure2}: here $U_{e,0}^\mathrm{G}=+2.5$~meV. Other simulation parameters are as in Fig.~\ref{fig:SAW:figure2}.
\label{fig:SAW:figure4}}
\end{figure*}

Note, however, that in the quantum simulation the internal structure of the SIX, i.e., electron-hole correlations, plays a role in determining the exact amount of tunneled wave packet. This is illustrated in Fig.~\ref{fig:SAW:figure4} by the difference between exact (black) and RIX (red) calculations. The latter shows both a broader distribution of the tunneled wave packet, i.e., lower Fourier components, and a substantial reflected wave packet which follows the dynamics of the more energetic classical trajectories. This is consistent with results which we obtained recently for stationary scattering potentials: neglecting electron-hole correlation in tunneling processes substantially underestimates the tunneling probability. This effect is induced by the mutual Coulomb attraction of the two carriers and was obtained in Ref.~\onlinecite{GrasselliPRB16b} for static potentials.
The point-like-particle (PLP) approximation, where the c.m.~wave packet is propagated under the potential computed at $x=0$, Eq.~(\ref{eq:pointlike}), gives results which are very similar to the RIX approximation, and it is not able to reproduce exact results, as it underestimates as well the tunneling probability. 
This is also shown in Fig.~\ref{fig:SAW:figure4}, reporting the simulations for a scattering intensity $U_{e,0}^\mathrm{G}$ that is similar to the SAW amplitude. Here, the value of $\rho_\mathrm{cm}(X;t)$ in the initial SAW minimum is lower in PLP calculations (blue) with respect to the exact results (black).

Indeed, as discussed in Sec.~\ref{sec:SEapproach}, the electron-hole dynamics may be lumped into a self-energy which accounts for virtual transitions during scattering. 
Since we have seen that the internal correlations do in fact influence the scattering of SAW-driven SIX if the potential is of the order of the SAW amplitude, it is interesting to ascertain whether the SE method gives results that are quantitatively correct when compared with the exact calculation for the present TD problem. 
Indeed, Fig.~\ref{fig:SAW:figure4} shows that the specific features of the exact $\rho_\mathrm{cm}(X;t)$ can be obtained very well within the self-energy approach, both qualitatively and quantitatively.

In Fig.~\ref{fig:SAW:figure2}(b), with the amplitude of the impurity potential that is twice the SAW amplitude, the classical distribution is very similar to the previous cases, while in the quantum simulation only a part of the wave packet tunnels through the scattering potential and remains localized in the SAW minimum. A significant part is reflected to the right, almost reproducing the dynamics of the most energetic part of the classical distribution. Note, again, that the fraction of transmitted or reflected wave packet is largely determined by the electron-hole correlations. Indeed, RIX calculations (not shown) overestimate reflection by about 100\%, in agreement with Ref.~\onlinecite{GrasselliPRB16b}. Again, as in the case of Fig.~\ref{fig:SAW:figure4}, the behavior of PLP is analogous to the RIX approximation, while the self-energy approximation reproduces the exact results with high accuracy. Hence, we omit to show $\rho_\mathrm{cm}(X;t)$ for RIX, PLP and self-energy approximations, redirecting their results to a more quantitative analysis on the c.m. time evolution given in Sec.~\ref{sec:cmdynamics}.

All in all, while classical trajectory distributions are almost unaffected by the scattering potential intensity, the quantum dynamics is strongly sensitive to the scattering strength, due to tunneling. Furthermore, electron-hole internal correlations determine the amount and spectral composition of the transmitted and reflected wave packets.

\subsubsection{c.m. dynamics}\label{sec:cmdynamics}

\begin{figure}
\begin{center}
\includegraphics[width=0.9\columnwidth]{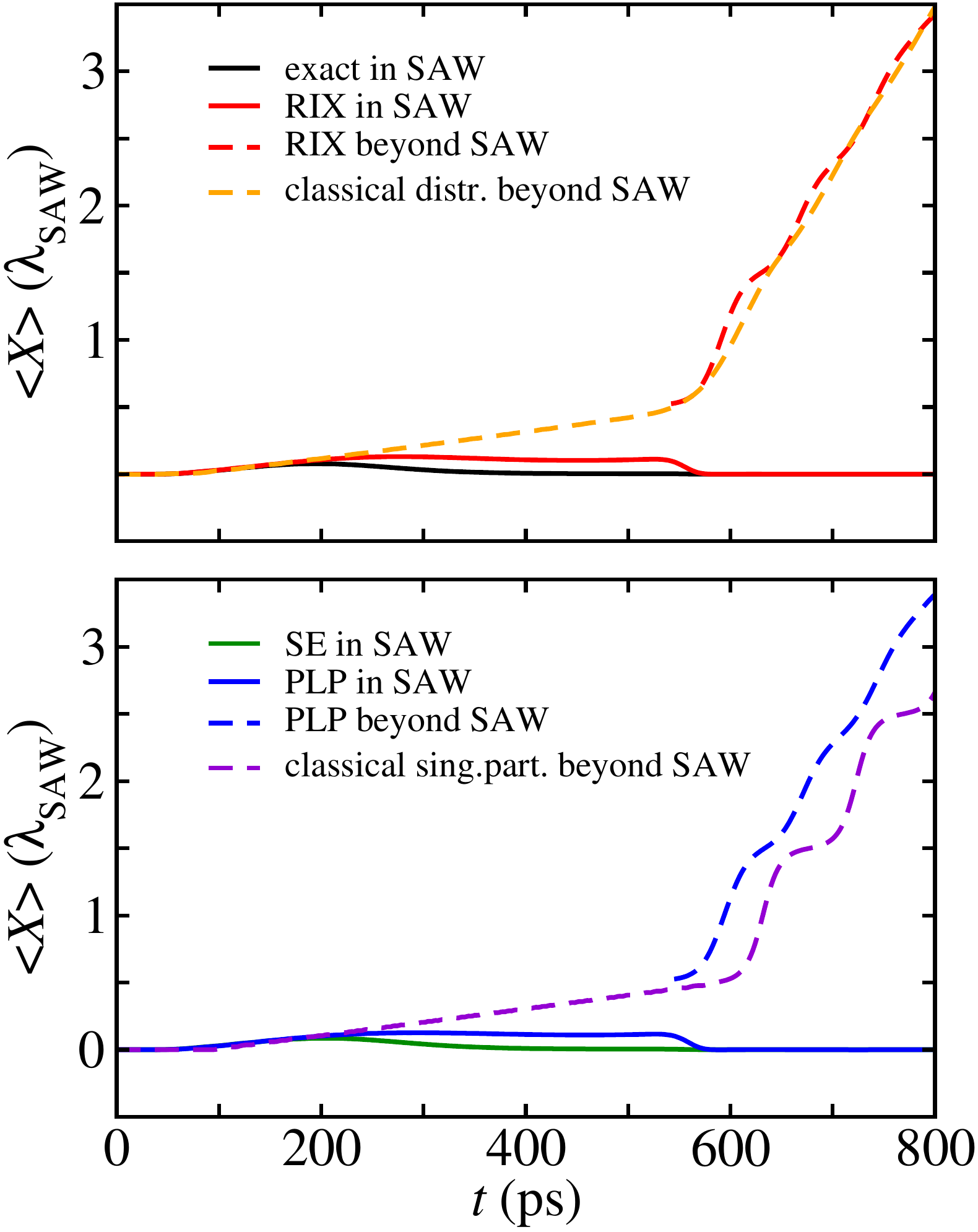}
\end{center}
\caption{\small $\langle X\rangle$(t) in the different models for the case of Fig.~\ref{fig:SAW:figure4} (intermediate scattering potential). Different curves are split among two panels for clarity.\label{fig:SAW:figure3}}
\end{figure}

\begin{figure}
\begin{center}
\includegraphics[width=0.9\columnwidth]{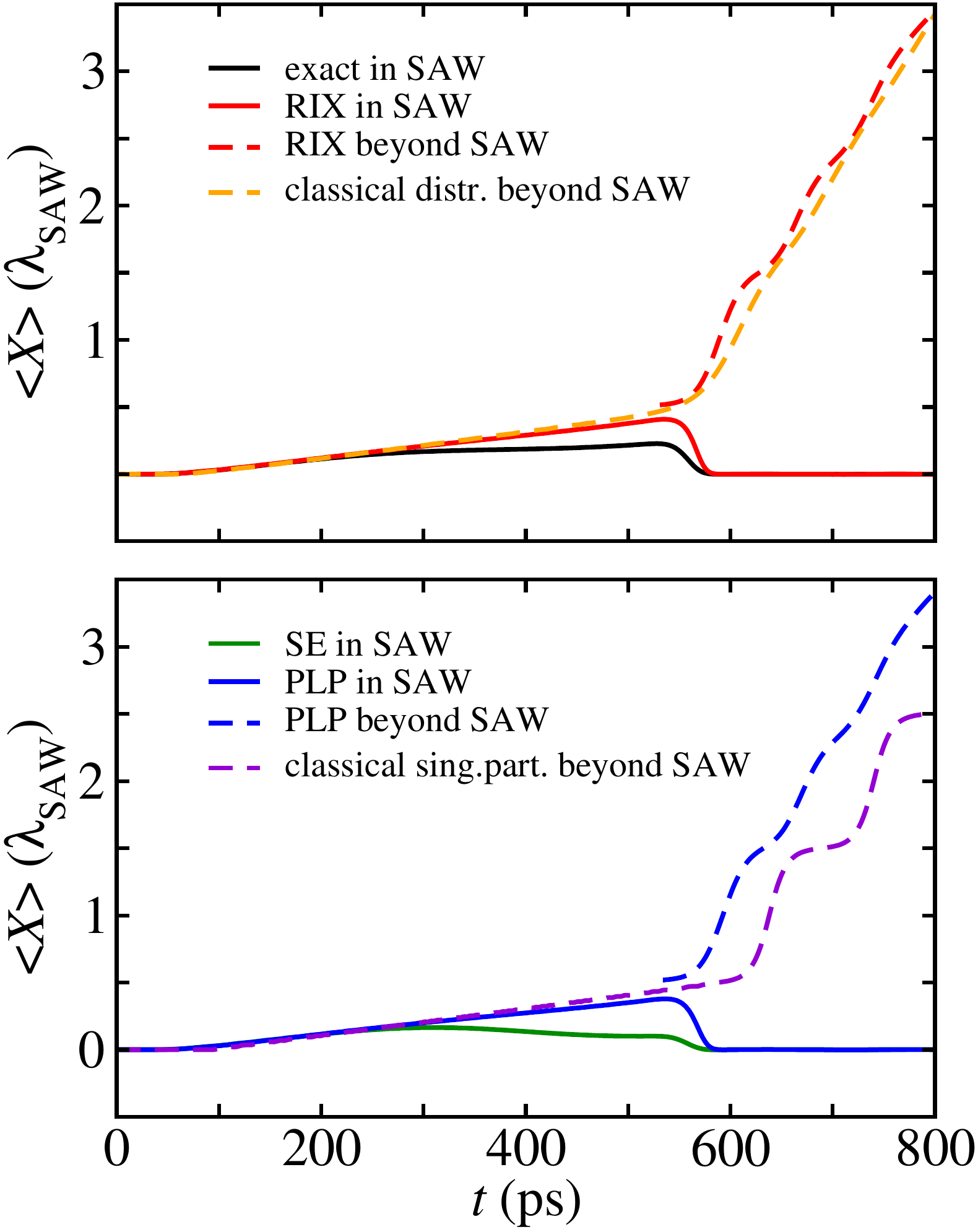}
\end{center}
\caption{\small $\langle X\rangle$(t) in the different models for the case of Fig.~\ref{fig:SAW:figure2}(b) (strong scattering potential). Different curves are split among two panels for clarity. \label{fig:SAW:figure3ter}}
\end{figure}

To make the analysis more quantitative, we compute the average c.m.~position as a function of time, $\langle X \rangle (t)$, using the different propagation methods, for the propagations shown in Fig.~\ref{fig:SAW:figure2}(b) and in Fig.~\ref{fig:SAW:figure4}.\footnote{{The quantity {{$\langle X \rangle (t)$}} is the expectation value for quantum simulations, while it is the average over the different trajectories for the classical evolution, as a function of time $t$.}}
Since the wave packet normally splits in two parts, we separately calculate the average in two regions, namely the SAW potential minimum where the wave packet is initialized (region A, $-\lambda_\mathrm{SAW}/2 < X < \lambda_\mathrm{SAW}/2$), and in the region of reflection (region B, $X > \lambda_\mathrm{SAW}/2$), namely the next three SAW minima in the direction opposite to the SAW propagation.
As expected, no part of the wave function is found in the opposite direction, i.e. the IX is never shot forward in the direction of the SAW propagation.
For clarity, we stress again that we study the dynamics of SIXs into the moving reference frame with the SAW at rest. As a consequence, the x axis is integral with the SAW potential during the dynamics shown in the figures. 

The corresponding averages, 
\begin{equation}
\langle X \rangle_\mathrm{A(B)} (t) = \frac{\int_\mathrm{A(B)} X \, \rho_\mathrm{c.m.}(X;t)\, dX}{\int_\mathrm{A(B)} \rho_\mathrm{c.m.}(X;t) \, dX}
\end{equation}
are shown in Figs.~\ref{fig:SAW:figure3} and \ref{fig:SAW:figure3ter}, with solid and dashed lines for regions A and B, respectively.\footnote{To avoid non-physical values we only accumulate the averages when $\rho_\mathrm{c.m.}(X;t) > \geq 5$\% of its total value. This explains, for instance, why the dashed line starts at an intermediate time and not from $t=0$.}

We start with the evolutions shown in Fig.~\ref{fig:SAW:figure4}, for which we report $\langle X \rangle_\mathrm{A}, \langle X \rangle_\mathrm{B} $ in Fig.~\ref{fig:SAW:figure3}. For clarity, we split different methods in two panels.
The SIX wave packet, initially with $\langle X \rangle_\mathrm{A} (t) = 0$  is displaced by the scattering barrier. 
While the classical particle moves at constant speed $v_\mathrm{SAW}$, since it is squeezed between the SAW and the scattering potential, the quantum wave packet tunnels through the scattering potential, and $\langle X \rangle_\mathrm{A} (t)$ stabilizes at $X \approx 0.2\, \lambda_\mathrm{SAW}$. 
However, when the position of the Gaussian external potential coincides with the maximum of the SAW, the classical particles are released with an initial boost and travel at a constant average velocity determined by the initial distribution [see Appendix \ref{App:ClassPeriodMot}].

The propagation method strongly determines the \textit{fraction} of the wave packet which is reflected, but not much its shape; therefore, the expectation values $\langle X \rangle_\mathrm{B} (t)$ (dashed lines in Figs.~\ref{fig:SAW:figure3} and \ref{fig:SAW:figure3ter}) differ only slightly, though obtained via different quantum propagation methods which give very different probabilities $\int_\mathrm{B} \rho_\mathrm{cm}(X;t) dX$. 
Furthermore, they are well described by $\langle X \rangle_\mathrm{B} (t)$ computed from the classical distribution.
On the contrary, since the shape of $\rho_\mathrm{cm}(X;t)$ \textit{inside} the initial SAW potential well is strongly determined by the tunneling probability, the different propagation methods give different laws $\langle X \rangle (t)$: in particular, the larger the tunneling, the closer $\langle X \rangle$ is to $X=0$, i.e. the minimum of the SAW potential well.

After the SAW has completely passed the scattering potential, the transmitted quantum wave packet recovers the initial shape inside the initial SAW minimum, with $\langle X \rangle_\mathrm{A} (t) = 0$. This is true for both the exact and the RIX propagations. 
However, as noted above, in the latter case a small part of the wave packet does not tunnel and is reflected, moving at an almost constant speed comparable with the classical distribution result. 

A similar behavior is shown in Fig.~\ref{fig:SAW:figure3ter}, corresponding to the process shown in Fig.~\ref{fig:SAW:figure2}(b). Due to the much smaller transmitted fraction of the wave packet, however, the classical and quantum average position of the c.m.~are similar until the process is over. In this case, of course, there is a reflected part of the wave packet both in the full and RIX propagations. Its average position (but not its amplitude, as noted above) coincide in the two methods and with the classical trajectories.

In Sec.~\ref{sec:strongpots} we pointed out that the PLP approximation, comparable to the RIX model, gives unsatisfactory results on the c.m. density profile, $\rho_\mathrm{cm}(X;t)$, while the self-energy correction to the RIX approach is able to recover most of the features of the exact $\rho_\mathrm{cm}(X;t)$. 
This is validated also by the computation of the expectation value of the c.m.~position, $\langle X \rangle (t)$, as shown in Fig.~\ref{fig:SAW:figure3}, where the exact (black) and self-energy approach (green) lines (plotted on the two different panels for clarity) almost coincide. 
For even higher tunneling barrier [see Fig.~\ref{fig:SAW:figure3ter}, corresponding to the propagation of Fig.~\ref{fig:SAW:figure2}(b)] the self-energy correction is still in good agreement with the exact propagation, with only a \textit{minor overestimation} of the tunneling,\cite{GrasselliPRB16b} resulting in an average c.m.~position closer to the SAW minimum than in the exact calculation.  

For completeness, we show, in the lower panels of Figs.~\ref{fig:SAW:figure3} and \ref{fig:SAW:figure3ter}, the classical motion $X(t)$ of a single SIX, initialized at rest ($\dot{X}(t) =0$) in the minimum of the initial SAW potential well ($X(0) = 0$). The qualitative phenomenology is reproduced, the particle accelerating in the regions closed to the SAW minima, and slowing nearby the crests; nevertheless, due to strong dependence on the initial conditions of the classical dynamics, and in particular of the period needed by the classical particle to travel from one SAW crest to the following one (see Appendix~\ref{App:ClassPeriodMot}), this one-particle classical law of motion quantitatively differs from the time-law obtained by averaging on the classical distribution of trajectories (orange lines in Figs.~\ref{fig:SAW:figure3} and \ref{fig:SAW:figure3ter}).

\subsection{Dependence on SAW amplitude}

\begin{figure*}
\includegraphics[width=\textwidth]{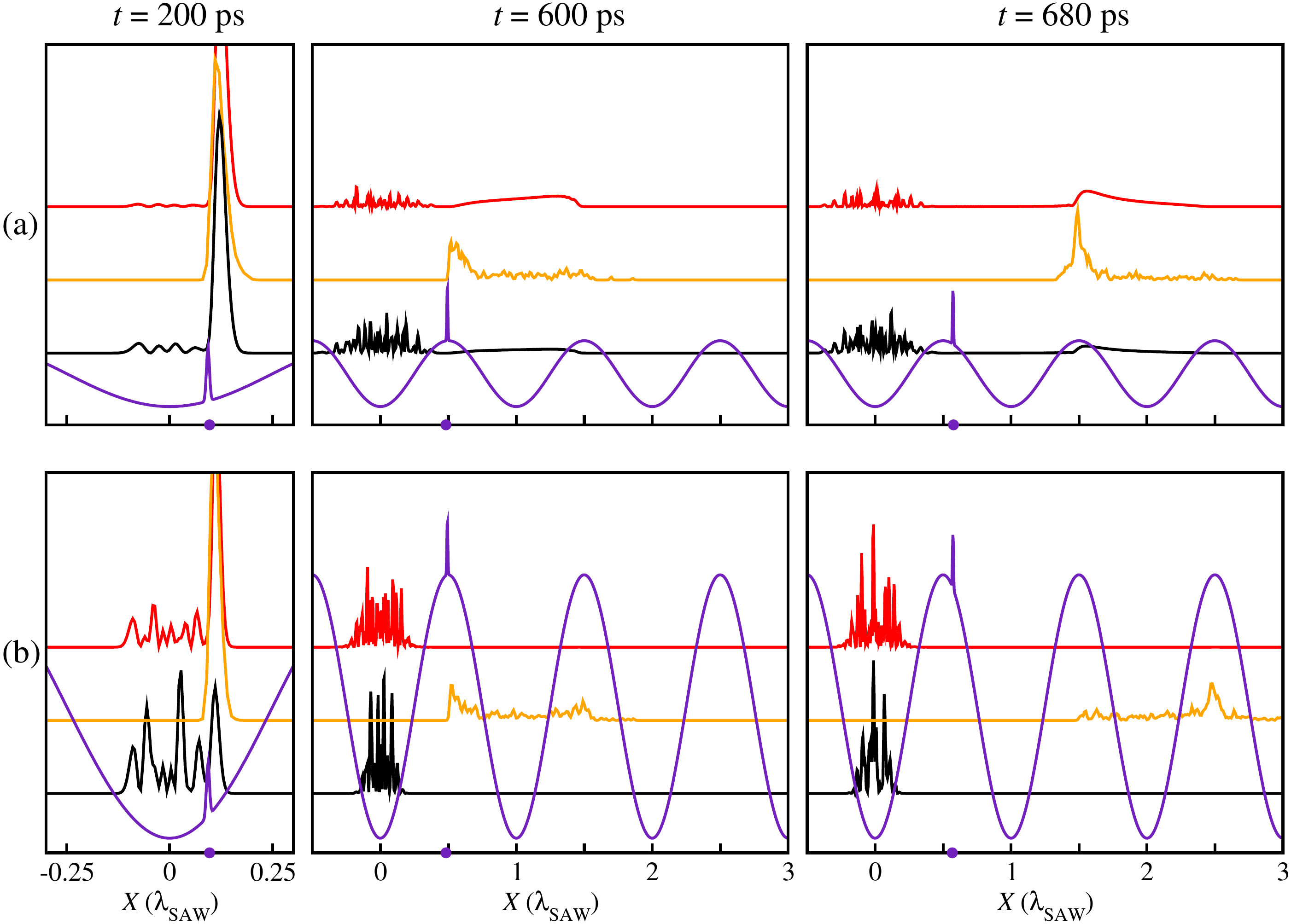}
\caption{Snapshots of simulation, taken at different times, for different total SAW amplitude. $U_{e,0} + U_{h,0} =$0.9 meV (a) and 3.6 meV (b). The PLP total potential is displayed in violet solid line. A violet dot on the $X$ axis indicates the position of the maximum of the Gaussian external potential. Exact (black line), RIX (red line), and classical (orange line) distributions are shown. \label{fig:SAW:figure_diff_ampl_orizz}}
\end{figure*}

We also performed simulations at different SAW amplitudes. 
The parameters of the Gaussian scattering potential are those of Fig.~\ref{fig:SAW:figure4}, i.e. $U_{e,0}^\mathrm{G}= +2.5\, \mathrm{meV}$, $U_{h,0}^\mathrm{G}= -1.0\, \mathrm{meV}$, and $\sigma_\mathrm{G}=10\,\mathrm{nm}$. 
Figure \ref{fig:SAW:figure_diff_ampl_orizz} shows snapshots of simulations, taken at different times, where the total SAW amplitude is (a) 0.9 meV and (b) 3.6 meV.\footnote{An intermediate regime, with SAW amplitude of 1.8 meV corresponds to the simulations already presented in Fig.~\ref{fig:SAW:figure4}, even if the vertical axis is rescaled, to allow the full representation of the potential in Fig.~\ref{fig:SAW:figure_diff_ampl_orizz}(b), keeping the same scale in the three panels.}

From the simulations, it is clear that, by increasing the SAW amplitude, the reflection speed of the SIX is also enhanced, as the SIX can convert into kinetic energy a greater amount of accumulated potential energy.
For instance, in Fig.~\ref{fig:SAW:figure_diff_ampl_orizz}(b) at $t=680\,\mathrm{ps}$ the classical SIX c.m.~distribution has overtaken the corresponding distribution in Fig.~\ref{fig:SAW:figure_diff_ampl_orizz}(a). Nevertheless, one has to keep in mind that, in the quantum case, almost the entire wave packet undergoes tunneling and remains confined to the original SAW minimum. 

\subsection{Broad scattering potentials}

Next we analyze the case of a broader scattering Gaussian potentials which is intended to mimic, e.g., shallow but extended potentials induced by dislocations or external gating. We fix the strength of the potential $U_{h,0}^\mathrm{G}=U_{e,0}^\mathrm{G}=+2.0$~meV and vary $\sigma_G$. 

\begin{figure*}
\includegraphics[width=\textwidth]{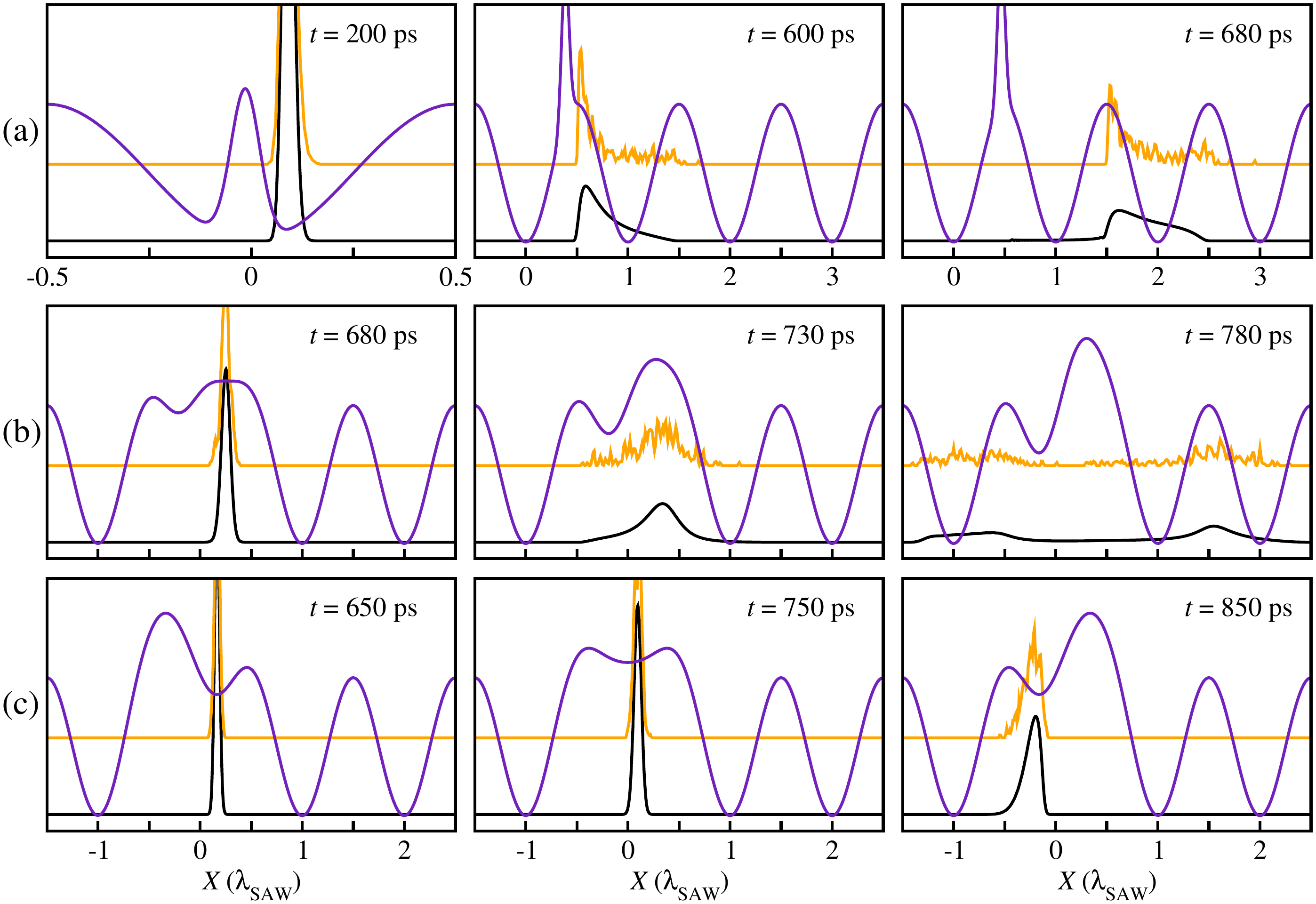}
\caption{Propagation in the case of smoother and stronger moving Gaussian. (a) $\sigma_G=100$~nm;  (b) $\sigma_G=600$~nm; (c) $\sigma_G=700$~nm. Exact quantum calculations (black) and classical simulations (orange) are shown. $U_{e,0}^\mathrm{G}=U_{h,0}^\mathrm{G}=+2.0$~meV in all panels. \label{fig:SAW:figureLargeSTD}}
\end{figure*}

Figure \ref{fig:SAW:figureLargeSTD} shows snapshots of the propagation at selected times with increasing $\sigma_\mathrm{G}$ from top to bottom. 
In all cases, different quantum propagation methods almost coincide and we only show $\rho_\mathrm{cm}(X;t)$ calculated from the exact propagation. This emphasizes that in smoother external potentials the c.m.~and relative motion DOFs are essentially uncoupled. Moreover, purely quantum phenomena are less important due to the vanishing tunneling probability in such extended potential barriers, and also the classical distribution is in good agreement with the quantum wave packet evolution. 

On the other hand, simulations suggest that the exciton dynamics is strongly dependent on the scattering potential extension, and slight variations imply very different evolution. Ultimately, this is due to the comparable size of the scattering potential width and the SAW wavelength.

In Fig.~\ref{fig:SAW:figureLargeSTD}(a) we show results for $\sigma_G = 100$~nm. In this case the SIX wave packet is completely reflected. However, when $\sigma_G \approx \lambda_\mathrm{SAW}$, the dynamics may become more complex. Figure \ref{fig:SAW:figureLargeSTD}(b) ($\sigma_G =600$~nm), shows the SIX distribution split in two parts, one reflected as in the $\sigma_G = 100$~nm case, and another one which is boosted forward: the latter distribution is now generated thanks to a plateau of the \textit{total} potential energy which forms during scattering because the SAW and scattering potentials are comparable both in strength and in width. Hence, part of the SIX distribution can also be pushed forward, after acquiring a potential energy equal to the double amplitude of the SAW. 

Surprisingly, for a somewhat larger standard deviation, $\sigma_G = \lambda_\mathrm{SAW}/4 =700$~nm (Fig.~\ref{fig:SAW:figureLargeSTD}(c)), the SIX distribution has a single peak in the initial SAW minimum: even if slightly excited, the c.m.~distribution is substantially not affected by the presence of the external potential. It should be emphasized that this case differs from the localization in the initial SAW well shown in Figs.~\ref{fig:SAW:figure2},\ref{fig:SAW:figure3}, and \ref{fig:SAW:figure3ter}). In that case (narrow potential) tunneling is possible, hence the different behavior of classical and quantum simulations. Here no tunneling is involved, and localization -- shown in both quantum and classical distribution -- is due to the specific time behavior of the total potential. In conclusion, the exciton propagation under broad potentials can be fairly well described classically. However, small variations in the external potential shape, as well as on the initial conditions, may substantially change the exciton dynamics: when simulating realistic devices, a detailed knowledge of the potential landscape and device parameters is likely to be necessary for qualitative correct predictions. Vice versa, the strong dependence of the SIX dynamics on the potential landscape may be exploited to get information on the latter, by adopting SAW-SIXs as fine probes.

\subsection{SAW-SIX devices}

\begin{figure}
\includegraphics[width=\columnwidth]{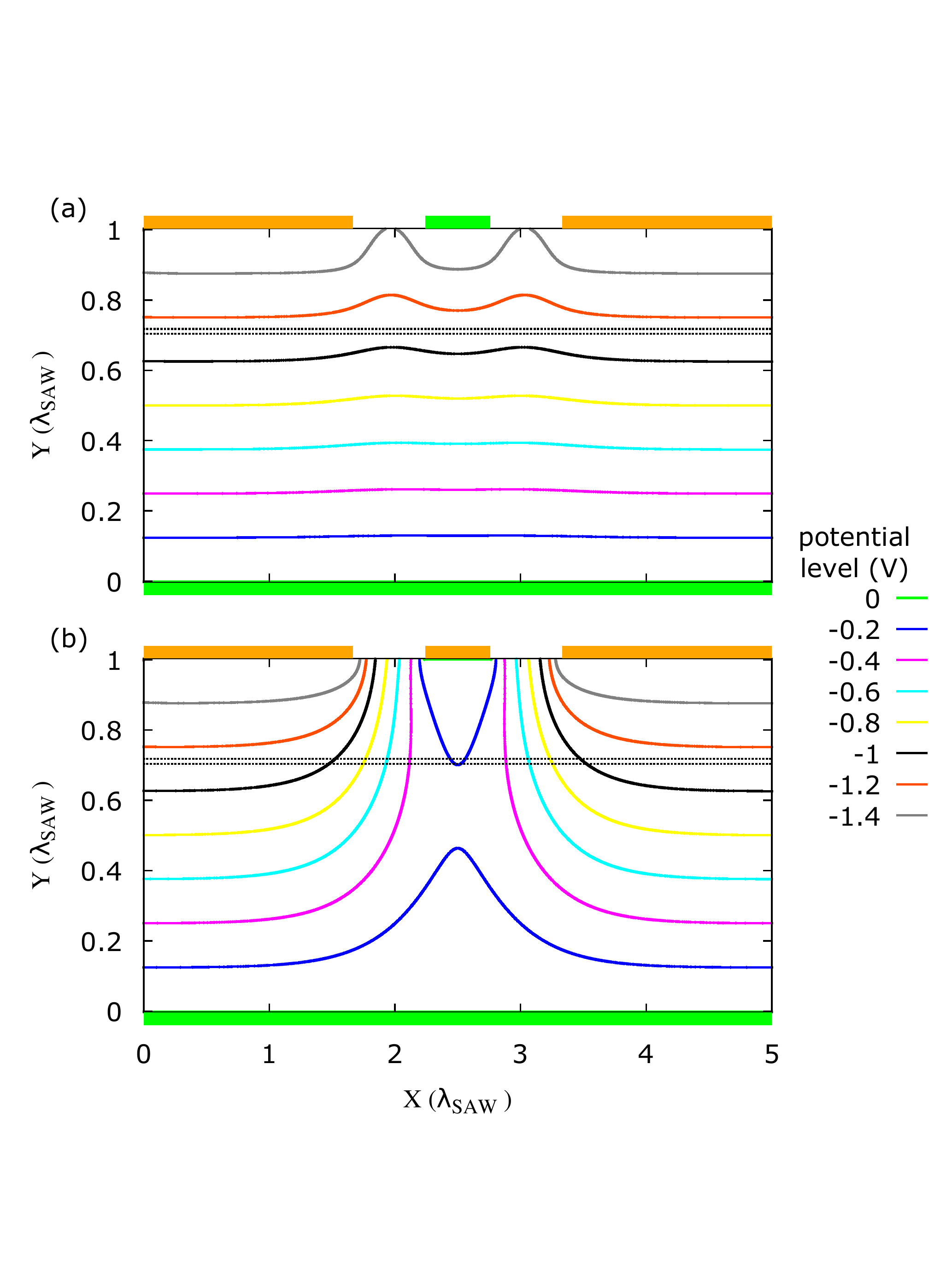}
\caption{Isopotential lines generated in a transverse section of the SAW-SIXs device by the back gate electrode (always at $0$~V) and by three top electrodes ($V_S$, $V_G$, $V_D$). 
(a) ON configuration, with $V_S = V_D = V_G =-1.6$~V. 
(b) OFF configuration, with $V_S = V_D = -1.6$~V, and $V_G=0$~V. 
The abscissa is the propagation direction $X$ while the $Y$ direction is orthogonal to the plane of the CQW, represented by the two dotted horizontal lines. 
\label{fig8:isopot2D}
}
\end{figure}
SAW-SIX devices have been investigated in Ref.~\onlinecite{ViolanteNJP14}. To investigate the dominant physical effects taking place in such experiments, we performed simulations with device parameters after Ref.~\onlinecite{ViolanteNJP14}. SIXs are propagated in an asymmetric 14-4-17~nm GaAs/Al$_{0.3}$Ga$_{0.7}$As/GaAs CQW heterostructure, located 800~nm under the top of the sample, whose total thickness is about one $\lambda_\mathrm{SAW}$.
The SAW parameters are $\lambda_\mathrm{SAW}=2.8\,\mathrm{\mu m}$, $v_\mathrm{SAW} = 2.5 \times 10^3$~m/s, and potential energy amplitude is +0.9~meV for both the electron and the hole.

The external gating on top of the device consists of three electrodes, termed source, gate and drain, of potentials $V_S$, $V_G$, and $V_D$, respectively. 
The gate electrode, 1.5~$\mathrm{\mu m}$ wide, is separated by the same distance from the source and the drain.
For the simulation, the electron-hole separation has been chosen as the distance between the mid planes of the two quantum wells, namely $d=19.5\,\mathrm{nm}$. 
To obtain the external potential landscape, we numerically solved the Laplace equation in the OFF configuration, where $V_S = V_D = -1.6$~V, and $V_G=0.0$~V, and in the ON configuration, where $V_S = V_D = V_G =-1.6$~V. 
Panels (a) and (b) in Fig.~\ref{fig8:isopot2D} show the isolines of the potential in the ON and OFF configurations, respectively. 
In the OFF configuration the SIX, initially generated under the source electrode, is driven by the SAW and encounters the gate barrier.
On the contrary, in the ON configuration the SIX dynamics is overall free, apart from the side-field potential which generates in the region between the gate and the source, and between the gate and the drain, electrodes.

\begin{figure}
\includegraphics[trim={0 0 0 8cm},clip,width=0.9\columnwidth]{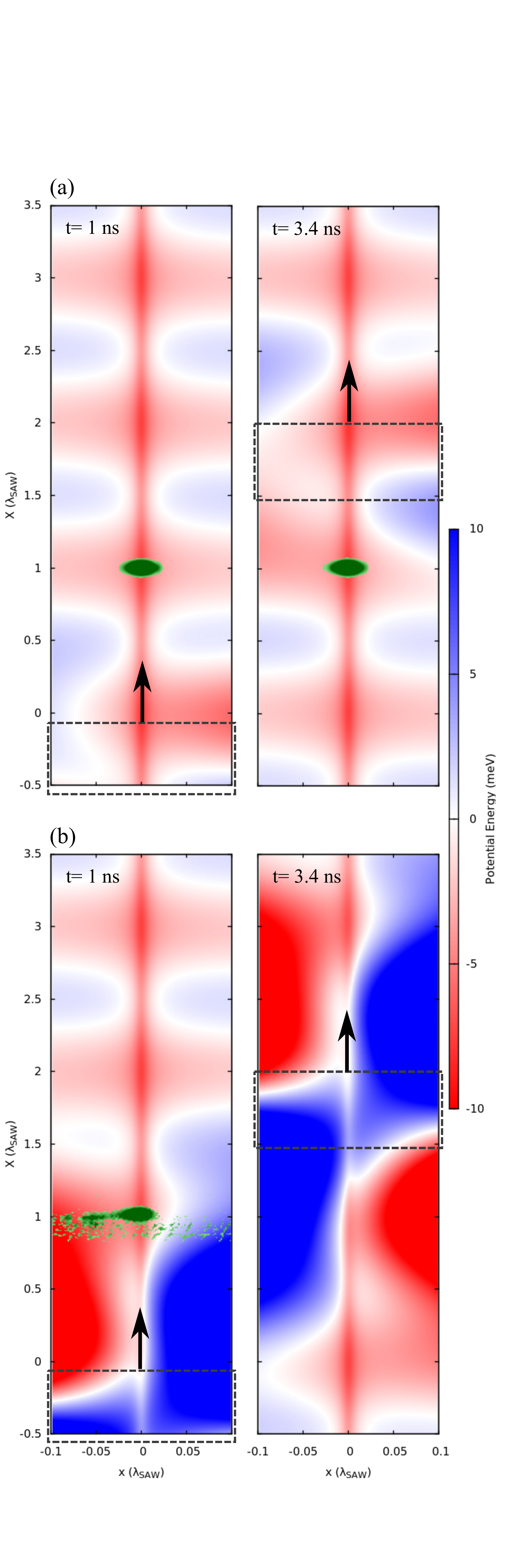}
\caption{$X-x$ representation of a SAW-driven SIX at two different times during the propagation in the potential generated by the electrodes of Fig.~\ref{fig8:isopot2D}. The top gate electrode is represented by the dashed rectangle. (a) and (b) panels show the ON and OFF configurations, respectively. The red/blue colormap represents the potential energy experienced by the SIX. The SIX wave function is initialized around $X=1$, $x=0$ and is represented in green. In the OFF configuration the SIX is fully dissociated.
\label{fig9:4panels}
}
\end{figure}

Although the real device is a two-dimensional system, the potential landscape is mainly modulated along one direction. Therefore, we assume that the two spatial directions are separable,\cite{GrasselliPRB16} and we simulate an effectively 1D system, similarly to previous sections. 

Figure~\ref{fig9:4panels} shows the representation, in c.m.~and relative coordinates, of the SAW potential and of $|\Psi(X,x;t)|^2$, $1$~ns (left) and $3.4$~ns (right) after the initialization of the SIX. 
The vertical axis represents the c.m. coordinate and the horizontal axis the relative one. Since we use a reference frame in which the SAW modulations are at rest, the dashed rectangle representing the position of the top gate electrode moves in the direction indicated by the black arrow with velocity $v_{SAW}$.
Panels (a) and (b) correspond to the ON and OFF configurations, respectively. 

In the ON configuration, the potential landscape generated by the gate electrode is weaker than both the confining SAW and the Coulomb interaction (represented by the vertical negative potential stripe around $x=0$). Therefore, although it pulls the electron and hole in opposite directions (horizontally in Fig.~\ref{fig9:4panels}), the SIX remains essentially unaltered at the bottom of the SAW during propagation. 
In the OFF configuration, at the contrary, the strong potential drop between the central gate and the source and drain lateral electrodes generates a strong in-plane potential gradient. Therefore, the dipolar force leads to  strong excitations of the internal modes of the electron-hole pair and, eventually, to SIX dissociation. 
This is shown by the strong potential modulation along the relative coordinate just before and after the gate electrode in panel (b). The negative potential overcomes the Coulomb attraction and the pair breaks. 
This effect starts in the left graph ($t=1$~ns) of panel (b), while in the right graph ($t=3.4$~ns) the SIX density is essentially zero in the represented domain, indicating complete ionization.

Note that in the ON configuration, the smooth external potentials varies on a length scale which is at least one order of magnitude larger than the effective Bohr radius $a_\mathrm{B}^*$. Therefore, separability between c.m.~and relative dynamics still holds. This is not the case in the OFF configuration, where strong potential gradients unavoidably transfers energy, gained from the SAW, from the c.m. to relative DOFs, eventually leading to dissociation. 

\section{Conclusions}\label{sec:Conclusions}

We simulated TD propagation of a coupled-quantum-well SIX driven by the potential generated by a non-piezoelectric SAW, scattering against different external potentials, for a wide range of scattering potential length scale, energy range, and smoothness. We exploited different propagation approaches, at different levels of approximation, from full quantum evolution of a Coulomb correlated $e$-$h$ pair to the Euler-Lagrange trajectories of classical point-like SIXs. 

We found a strong dependence of the SIX dynamics on details of the external potential landscape. 
For instance, localized potentials can temporarily stop the SIX transport guided by the SAW, and induce high-speed back reflection due to the conversion into c.m.~kinetic energy of the SAW potential energy amplitude. 
Shallow and extended scattering potentials allow for a classical description of the SAW-SIX dynamics, since the quantum correlation between the c.m.~and the relative motions of the $e$-$h$ pair, as well as tunneling processes, can be neglected.  

SAW-SIX dynamics in steep scattering potentials, on the contrary, needs a quantum description to include quantum tunneling and interference processes. Moreover, the coupling between the c.m.-relative DOFs is responsible for quantitative estimates of the scattering coefficients, and a global description of the SIX evolution. Besides the exact propagation of all the quantum DOFs, which is, unfortunately, computationally expensive, such correlation effects can be accurately included in the c.m.~evolution by means of a properly design local self-energy contribution, which takes into account virtual transitions to excited relative-motion levels. This self-energy term must be added to the effective potential seen by the SIX c.m. in the relative-coordinate mean field, which fails, by itself, to describe the c.m.~dynamics properly. 

An experimental detection of the detailed correlated dynamics of the SIX could be obtained via spatially- and energy-resolved photoluminescence spectroscopy, but the detection of SIX dissociation or its localization in specific SAW minima should be less challenging from the technological perspecctive, and sufficient to reveal the quantum effects addressed in this work.

Simulations performed in experimentally attainable devices, whose parameters are taken from the literature, showed non-vanishing dissociation probability of SIX due to the strength of the potential landscape  and side fields generated by external gating. This may have a profound impact on the efficiency of realistic opto-excitonic devices\cite{Butov_SLAMS17}, as the photo-luminescence emission due to $e$-$h$ recombination is prevented by exciton dissociation.

\begin{acknowledgments}
F.G. acknowledges L. Butov and M. Fogler for useful discussions. We are grateful to P. Bordone for fruitful discussions and for suggesting several SAW-related references. We acknowledge the CINECA award under Iscra C project IsC33-“FUQUDIX.” for computing time on HPC architectures.
\end{acknowledgments}

\begin{appendix}

\section{$X-x$ representation}\label{App:Xxrepresentation}

\begin{figure*}
\includegraphics[width=\textwidth]{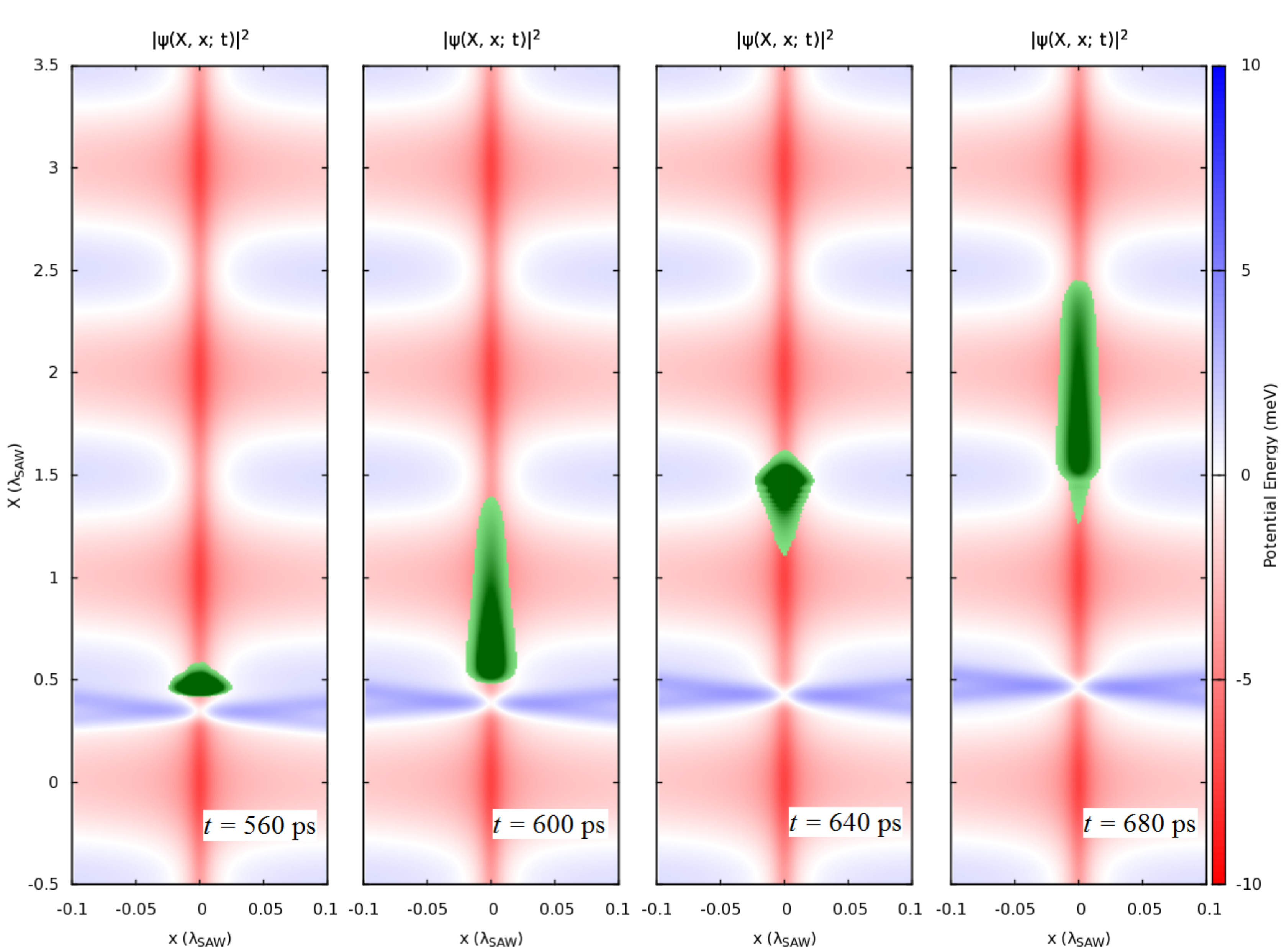}
\caption{Propagation in the case of smoother ($\sigma_G=100$~nm) and stronger ($U_{e,0}^\mathrm{G}=U_{h,0}^\mathrm{G}=+2.0$~meV) moving Gaussian in the full $X-x$ representation.\label{fig:SAW:figureA1}}
\end{figure*}

For clarity, we show in Fig.~\ref{fig:SAW:figureA1} some snapshot of the simulation already discussed in Fig.~\ref{fig:SAW:figureLargeSTD}(a) concerning the exact propagation, where both the c.m.~and the relative degrees of freedom are displayed. In green, we show the full 2-DOFs probability density, $|\Psi(X,x;t)|^2$, from which the marginal probability $\rho_\mathrm{cm}(X;t) \equiv \int |\Psi(X,x;t)|^2 dx$ of Fig.~\ref{fig:SAW:figureLargeSTD}(a) (black line) is computed after integrating on the relative coordinate. In red and blue color map, the total potential landscape is displayed: the vertical red stripe centered at $x=0$ is the electron-hole Coulomb interaction; the periodic sinusoidal part is the SAW potential: even if the SAW separately acts on the electron and the hole, it is so smooth that it is almost invariant along the relative coordinate, and therefore no c.m.~and relative motions coupling subsists due to the SAW. Proof of this is the fact that the relative part of the wave packet remains into the ground state of the unperturbed free IX, $\phi_0(x)$. The external Gaussian potential is the time dependent part of the potential landscape. 

\section{Classical particle periodic motion}\label{App:ClassPeriodMot}
\begin{figure}
\begin{center}
\includegraphics[width=0.9\columnwidth]{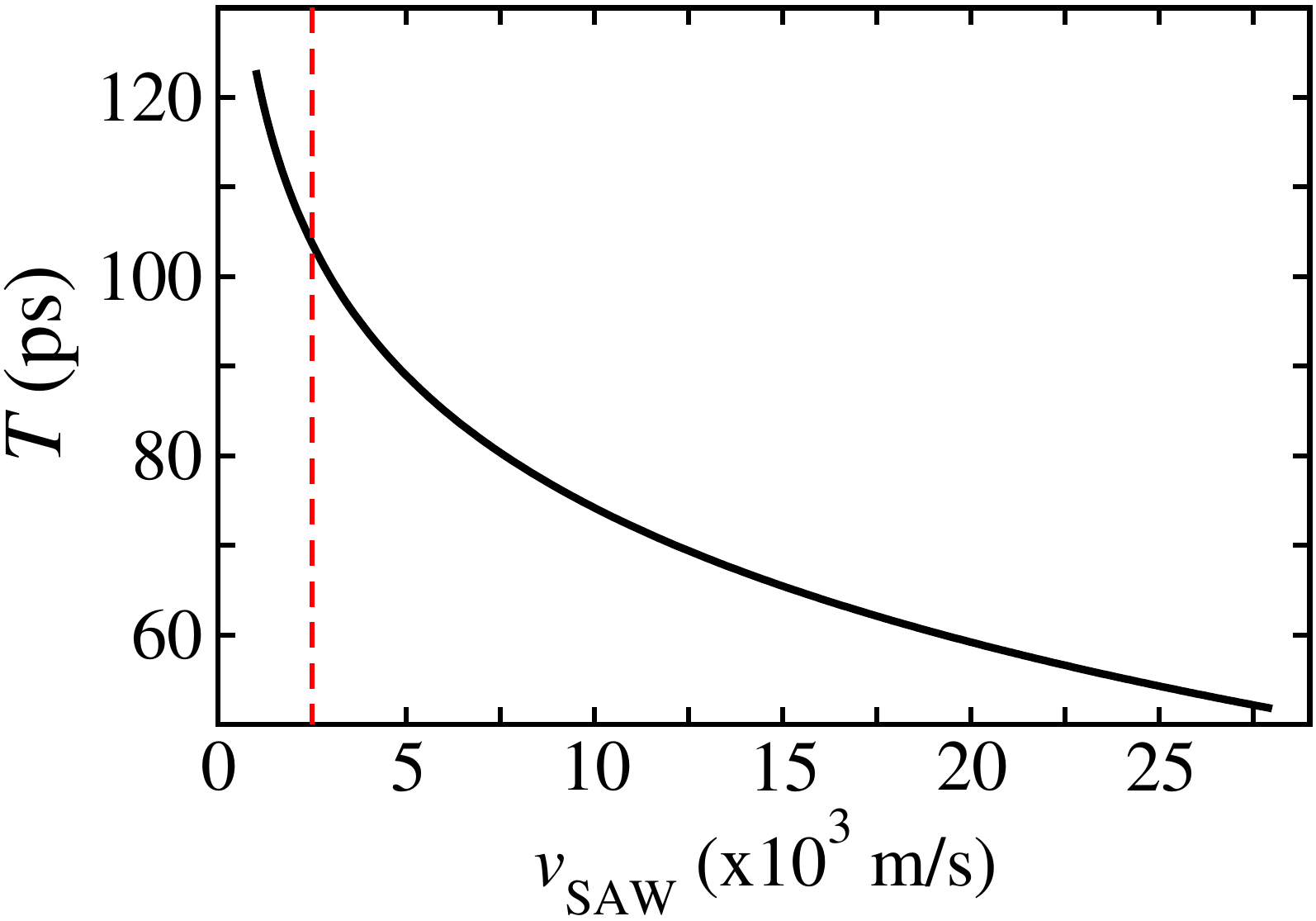}
\end{center}
\caption{Behavior of $T=T(v_\mathrm{SAW})$ (black solid line) for the parameters of the simulations. The vertical dashed red line indicates the SAW velocity as measured in Ref.~\cite{ViolanteNJP14}, $v_\mathrm{SAW} = 2.5\times 10^3$~m/s.\label{fig:SAW:figureA2}}
\end{figure}

Consider a single classical particle on the top of a crest of a cosinusoidal potential energy profile $U_0^\mathrm{SAW} \cos(2\pi X/\lambda_\mathrm{SAW})$ with spatial period $\lambda_\mathrm{SAW}$, at time $t^\ast$, pushed with initial velocity $\dot{X}(t^\ast)=V$. For $t>t^\ast$, the particle separates from the external pushing force, and will no longer feel it. The motion is therefore energy--conservative, and we can write
\begin{equation}
\frac{1}{2}M V^2 + U_0^\mathrm{SAW} = \frac{1}{2}M \dot{X}^2 + U_0^\mathrm{SAW} \cos(2\pi X/\lambda_\mathrm{SAW})
\end{equation}        
which can be integrated to give the period to travel one wave length $\lambda_\mathrm{SAW}$
\begin{equation}
T=\int_0^{\lambda_\mathrm{SAW}} \left[V^2 + \frac{4}{M}U_0^\mathrm{SAW}\sin^2(2\pi X/\lambda_\mathrm{SAW})\right]^{-1/2}.
\end{equation}
This integral can be calculated via the incomplete elliptic integral of the first kind $F(\varphi,u)$ of modulus $u$ and parameter $\varphi$  
\begin{equation}
\begin{split}
T &= \frac{\lambda_\mathrm{SAW}}{2\pi V} \int_0^{2\pi} \frac{d\theta}{\sqrt{1+q^2\sin^2(\theta)}} =\frac{\lambda_\mathrm{SAW}}{2\pi V} F\left(2\pi,iq\right) \\
&= 4\frac{\lambda_\mathrm{SAW}}{2\pi V} K(iq)
\end{split}
\end{equation}
where $K(u)\equiv F(\pi/2,u)$ is the complete elliptic integral of the first kind of modulus $u$, and $q^2\equiv 4U_0^\mathrm{SAW}/(MV^2) > 0$. Figure~\ref{fig:SAW:figureA2} shows $T=T(v_\mathrm{SAW})$, obtained by substituting the SAW parameters $U_0^\mathrm{SAW}=1.8$~meV, $\lambda_\mathrm{SAW}=2.8\,\mu$m, $V=v_\mathrm{SAW}$, and using the Wolfram's-Mathematica--tabulated values for the elliptic complete integrals.

\end{appendix}


\begin{thebibliography}{36}%
\makeatletter
\providecommand \@ifxundefined [1]{%
 \@ifx{#1\undefined}
}%
\providecommand \@ifnum [1]{%
 \ifnum #1\expandafter \@firstoftwo
 \else \expandafter \@secondoftwo
 \fi
}%
\providecommand \@ifx [1]{%
 \ifx #1\expandafter \@firstoftwo
 \else \expandafter \@secondoftwo
 \fi
}%
\providecommand \natexlab [1]{#1}%
\providecommand \enquote  [1]{``#1''}%
\providecommand \bibnamefont  [1]{#1}%
\providecommand \bibfnamefont [1]{#1}%
\providecommand \citenamefont [1]{#1}%
\providecommand \href@noop [0]{\@secondoftwo}%
\providecommand \href [0]{\begingroup \@sanitize@url \@href}%
\providecommand \@href[1]{\@@startlink{#1}\@@href}%
\providecommand \@@href[1]{\endgroup#1\@@endlink}%
\providecommand \@sanitize@url [0]{\catcode `\\12\catcode `\$12\catcode
  `\&12\catcode `\#12\catcode `\^12\catcode `\_12\catcode `\%12\relax}%
\providecommand \@@startlink[1]{}%
\providecommand \@@endlink[0]{}%
\providecommand \url  [0]{\begingroup\@sanitize@url \@url }%
\providecommand \@url [1]{\endgroup\@href {#1}{\urlprefix }}%
\providecommand \urlprefix  [0]{URL }%
\providecommand \Eprint [0]{\href }%
\providecommand \doibase [0]{http://dx.doi.org/}%
\providecommand \selectlanguage [0]{\@gobble}%
\providecommand \bibinfo  [0]{\@secondoftwo}%
\providecommand \bibfield  [0]{\@secondoftwo}%
\providecommand \translation [1]{[#1]}%
\providecommand \BibitemOpen [0]{}%
\providecommand \bibitemStop [0]{}%
\providecommand \bibitemNoStop [0]{.\EOS\space}%
\providecommand \EOS [0]{\spacefactor3000\relax}%
\providecommand \BibitemShut  [1]{\csname bibitem#1\endcsname}%
\let\auto@bib@innerbib\@empty
\bibitem [{\citenamefont {Datta}(1986)}]{Datta_SAW86}%
  \BibitemOpen
  \bibfield  {author} {\bibinfo {author} {\bibfnamefont {S.}~\bibnamefont
  {Datta}},\ }\href@noop {} {\emph {\bibinfo {title} {{Surface acoustic wave
  devices}}}}\ (\bibinfo  {publisher} {Prentice-Hall},\ \bibinfo {address}
  {Englewood Cliffs, N.J., USA},\ \bibinfo {year} {1986})\BibitemShut {NoStop}%
\bibitem [{\citenamefont {Kovalev}\ and\ \citenamefont
  {Chaplik}(2015)}]{Kovalev_JETPL15}%
  \BibitemOpen
  \bibfield  {author} {\bibinfo {author} {\bibfnamefont {V.~M.}\ \bibnamefont
  {Kovalev}}\ and\ \bibinfo {author} {\bibfnamefont {A.~V.}\ \bibnamefont
  {Chaplik}},\ }\href {\doibase 10.1134/S002136401503008X} {\bibfield
  {journal} {\bibinfo  {journal} {JETP Letters}\ }\textbf {\bibinfo {volume}
  {101}},\ \bibinfo {pages} {177} (\bibinfo {year} {2015})}\BibitemShut
  {NoStop}%
\bibitem [{\citenamefont {Wei\ss{}}\ \emph {et~al.}(2018)\citenamefont
  {Wei\ss{}}, \citenamefont {H\"orner}, \citenamefont {Zallo}, \citenamefont
  {Atkinson}, \citenamefont {Rastelli}, \citenamefont {Schmidt}, \citenamefont
  {Wixforth},\ and\ \citenamefont {Krenner}}]{Weiss_PRApp18}%
  \BibitemOpen
  \bibfield  {author} {\bibinfo {author} {\bibfnamefont {M.}~\bibnamefont
  {Wei\ss{}}}, \bibinfo {author} {\bibfnamefont {A.~L.}\ \bibnamefont
  {H\"orner}}, \bibinfo {author} {\bibfnamefont {E.}~\bibnamefont {Zallo}},
  \bibinfo {author} {\bibfnamefont {P.}~\bibnamefont {Atkinson}}, \bibinfo
  {author} {\bibfnamefont {A.}~\bibnamefont {Rastelli}}, \bibinfo {author}
  {\bibfnamefont {O.~G.}\ \bibnamefont {Schmidt}}, \bibinfo {author}
  {\bibfnamefont {A.}~\bibnamefont {Wixforth}}, \ and\ \bibinfo {author}
  {\bibfnamefont {H.~J.}\ \bibnamefont {Krenner}},\ }\href {\doibase
  10.1103/PhysRevApplied.9.014004} {\bibfield  {journal} {\bibinfo  {journal}
  {Phys. Rev. Applied}\ }\textbf {\bibinfo {volume} {9}},\ \bibinfo {pages}
  {014004} (\bibinfo {year} {2018})}\BibitemShut {NoStop}%
\bibitem [{\citenamefont {Violante}\ \emph {et~al.}(2014)\citenamefont
  {Violante}, \citenamefont {Cohen}, \citenamefont {Lazić}, \citenamefont
  {Hey}, \citenamefont {Rapaport},\ and\ \citenamefont
  {Santos}}]{ViolanteNJP14}%
  \BibitemOpen
  \bibfield  {author} {\bibinfo {author} {\bibfnamefont {A.}~\bibnamefont
  {Violante}}, \bibinfo {author} {\bibfnamefont {K.}~\bibnamefont {Cohen}},
  \bibinfo {author} {\bibfnamefont {S.}~\bibnamefont {Lazić}}, \bibinfo
  {author} {\bibfnamefont {R.}~\bibnamefont {Hey}}, \bibinfo {author}
  {\bibfnamefont {R.}~\bibnamefont {Rapaport}}, \ and\ \bibinfo {author}
  {\bibfnamefont {P.~V.}\ \bibnamefont {Santos}},\ }\href
  {http://stacks.iop.org/1367-2630/16/i=3/a=033035} {\bibfield  {journal}
  {\bibinfo  {journal} {New Journal of Physics}\ }\textbf {\bibinfo {volume}
  {16}},\ \bibinfo {pages} {033035} (\bibinfo {year} {2014})}\BibitemShut
  {NoStop}%
\bibitem [{\citenamefont {High}\ \emph {et~al.}(2008)\citenamefont {High},
  \citenamefont {Novitskaya}, \citenamefont {Butov}, \citenamefont {Hanson},\
  and\ \citenamefont {Gossard}}]{High_Science11072008}%
  \BibitemOpen
  \bibfield  {author} {\bibinfo {author} {\bibfnamefont {A.~A.}\ \bibnamefont
  {High}}, \bibinfo {author} {\bibfnamefont {E.~E.}\ \bibnamefont
  {Novitskaya}}, \bibinfo {author} {\bibfnamefont {L.~V.}\ \bibnamefont
  {Butov}}, \bibinfo {author} {\bibfnamefont {M.}~\bibnamefont {Hanson}}, \
  and\ \bibinfo {author} {\bibfnamefont {A.~C.}\ \bibnamefont {Gossard}},\
  }\href {\doibase 10.1126/science.1157845} {\bibfield  {journal} {\bibinfo
  {journal} {Science}\ }\textbf {\bibinfo {volume} {321}},\ \bibinfo {pages}
  {229} (\bibinfo {year} {2008})},\ \Eprint
  {http://arxiv.org/abs/http://www.sciencemag.org/content/321/5886/229.full.pdf}
  {http://www.sciencemag.org/content/321/5886/229.full.pdf} \BibitemShut
  {NoStop}%
\bibitem [{\citenamefont {Sivalertporn}\ \emph {et~al.}(2012)\citenamefont
  {Sivalertporn}, \citenamefont {Mouchliadis}, \citenamefont {Ivanov},
  \citenamefont {Philp},\ and\ \citenamefont {Muljarov}}]{Sivalertporn_PRB12}%
  \BibitemOpen
  \bibfield  {author} {\bibinfo {author} {\bibfnamefont {K.}~\bibnamefont
  {Sivalertporn}}, \bibinfo {author} {\bibfnamefont {L.}~\bibnamefont
  {Mouchliadis}}, \bibinfo {author} {\bibfnamefont {A.~L.}\ \bibnamefont
  {Ivanov}}, \bibinfo {author} {\bibfnamefont {R.}~\bibnamefont {Philp}}, \
  and\ \bibinfo {author} {\bibfnamefont {E.~A.}\ \bibnamefont {Muljarov}},\
  }\href {\doibase 10.1103/PhysRevB.85.045207} {\bibfield  {journal} {\bibinfo
  {journal} {Phys. Rev. B}\ }\textbf {\bibinfo {volume} {85}},\ \bibinfo
  {pages} {045207} (\bibinfo {year} {2012})}\BibitemShut {NoStop}%
\bibitem [{\citenamefont {Andreakou}\ \emph {et~al.}(2015)\citenamefont
  {Andreakou}, \citenamefont {Cronenberger}, \citenamefont {Scalbert},
  \citenamefont {Nalitov}, \citenamefont {Gippius}, \citenamefont {Kavokin},
  \citenamefont {Nawrocki}, \citenamefont {Leonard}, \citenamefont {Butov},
  \citenamefont {Campman}, \citenamefont {Gossard},\ and\ \citenamefont
  {Vladimirova}}]{AndreakouPRB15}%
  \BibitemOpen
  \bibfield  {author} {\bibinfo {author} {\bibfnamefont {P.}~\bibnamefont
  {Andreakou}}, \bibinfo {author} {\bibfnamefont {S.}~\bibnamefont
  {Cronenberger}}, \bibinfo {author} {\bibfnamefont {D.}~\bibnamefont
  {Scalbert}}, \bibinfo {author} {\bibfnamefont {A.}~\bibnamefont {Nalitov}},
  \bibinfo {author} {\bibfnamefont {N.~A.}\ \bibnamefont {Gippius}}, \bibinfo
  {author} {\bibfnamefont {A.~V.}\ \bibnamefont {Kavokin}}, \bibinfo {author}
  {\bibfnamefont {M.}~\bibnamefont {Nawrocki}}, \bibinfo {author}
  {\bibfnamefont {J.~R.}\ \bibnamefont {Leonard}}, \bibinfo {author}
  {\bibfnamefont {L.~V.}\ \bibnamefont {Butov}}, \bibinfo {author}
  {\bibfnamefont {K.~L.}\ \bibnamefont {Campman}}, \bibinfo {author}
  {\bibfnamefont {A.~C.}\ \bibnamefont {Gossard}}, \ and\ \bibinfo {author}
  {\bibfnamefont {M.}~\bibnamefont {Vladimirova}},\ }\href {\doibase
  10.1103/PhysRevB.91.125437} {\bibfield  {journal} {\bibinfo  {journal} {Phys.
  Rev. B}\ }\textbf {\bibinfo {volume} {91}},\ \bibinfo {pages} {125437}
  (\bibinfo {year} {2015})}\BibitemShut {NoStop}%
\bibitem [{\citenamefont {Kovalev}\ and\ \citenamefont
  {Chaplik}(2016)}]{Kovalev_JETP16}%
  \BibitemOpen
  \bibfield  {author} {\bibinfo {author} {\bibfnamefont {V.~M.}\ \bibnamefont
  {Kovalev}}\ and\ \bibinfo {author} {\bibfnamefont {A.~V.}\ \bibnamefont
  {Chaplik}},\ }\href {\doibase 10.1134/S1063776116030158} {\bibfield
  {journal} {\bibinfo  {journal} {Journal of Experimental and Theoretical
  Physics}\ }\textbf {\bibinfo {volume} {122}},\ \bibinfo {pages} {499}
  (\bibinfo {year} {2016})}\BibitemShut {NoStop}%
\bibitem [{\citenamefont {Beian}\ \emph {et~al.}(2017)\citenamefont {Beian},
  \citenamefont {Dang}, \citenamefont {Alloing}, \citenamefont {Anankine},
  \citenamefont {Cambril}, \citenamefont {Gomez}, \citenamefont {Osmond},
  \citenamefont {Lema\^{\i}tre},\ and\ \citenamefont {Dubin}}]{BeianPRAPP17}%
  \BibitemOpen
  \bibfield  {author} {\bibinfo {author} {\bibfnamefont {M.}~\bibnamefont
  {Beian}}, \bibinfo {author} {\bibfnamefont {S.}~\bibnamefont {Dang}},
  \bibinfo {author} {\bibfnamefont {M.}~\bibnamefont {Alloing}}, \bibinfo
  {author} {\bibfnamefont {R.}~\bibnamefont {Anankine}}, \bibinfo {author}
  {\bibfnamefont {E.}~\bibnamefont {Cambril}}, \bibinfo {author} {\bibfnamefont
  {C.}~\bibnamefont {Gomez}}, \bibinfo {author} {\bibfnamefont
  {J.}~\bibnamefont {Osmond}}, \bibinfo {author} {\bibfnamefont
  {A.}~\bibnamefont {Lema\^{\i}tre}}, \ and\ \bibinfo {author} {\bibfnamefont
  {F.}\ \bibnamefont {Dubin}},\ }\href {\doibase
  10.1103/PhysRevApplied.8.054025} {\bibfield  {journal} {\bibinfo  {journal}
  {Phys. Rev. Applied}\ }\textbf {\bibinfo {volume} {8}},\ \bibinfo {pages}
  {054025} (\bibinfo {year} {2017})}\BibitemShut {NoStop}%
\bibitem [{\citenamefont {Wilkes}\ and\ \citenamefont
  {Muljarov}(2016)}]{WilkesPRB16}%
  \BibitemOpen
  \bibfield  {author} {\bibinfo {author} {\bibfnamefont {J.}~\bibnamefont
  {Wilkes}}\ and\ \bibinfo {author} {\bibfnamefont {E.~A.}\ \bibnamefont
  {Muljarov}},\ }\href {\doibase 10.1103/PhysRevB.94.125310} {\bibfield
  {journal} {\bibinfo  {journal} {Phys. Rev. B}\ }\textbf {\bibinfo {volume}
  {94}},\ \bibinfo {pages} {125310} (\bibinfo {year} {2016})}\BibitemShut
  {NoStop}%
\bibitem [{\citenamefont {Kuznetsova}\ \emph {et~al.}(2017)\citenamefont
  {Kuznetsova}, \citenamefont {Dorow}, \citenamefont {Calman}, \citenamefont
  {Butov}, \citenamefont {Wilkes}, \citenamefont {Muljarov}, \citenamefont
  {Campman},\ and\ \citenamefont {Gossard}}]{Kuznetsova_PRB17}%
  \BibitemOpen
  \bibfield  {author} {\bibinfo {author} {\bibfnamefont {Y.~Y.}\ \bibnamefont
  {Kuznetsova}}, \bibinfo {author} {\bibfnamefont {C.~J.}\ \bibnamefont
  {Dorow}}, \bibinfo {author} {\bibfnamefont {E.~V.}\ \bibnamefont {Calman}},
  \bibinfo {author} {\bibfnamefont {L.~V.}\ \bibnamefont {Butov}}, \bibinfo
  {author} {\bibfnamefont {J.}~\bibnamefont {Wilkes}}, \bibinfo {author}
  {\bibfnamefont {E.~A.}\ \bibnamefont {Muljarov}}, \bibinfo {author}
  {\bibfnamefont {K.~L.}\ \bibnamefont {Campman}}, \ and\ \bibinfo {author}
  {\bibfnamefont {A.~C.}\ \bibnamefont {Gossard}},\ }\href {\doibase
  10.1103/PhysRevB.95.125304} {\bibfield  {journal} {\bibinfo  {journal} {Phys.
  Rev. B}\ }\textbf {\bibinfo {volume} {95}},\ \bibinfo {pages} {125304}
  (\bibinfo {year} {2017})}\BibitemShut {NoStop}%
\bibitem [{\citenamefont {Astley}\ \emph {et~al.}(2007)\citenamefont {Astley},
  \citenamefont {Kataoka}, \citenamefont {Ford}, \citenamefont {Barnes},
  \citenamefont {Anderson}, \citenamefont {Jones}, \citenamefont {Farrer},
  \citenamefont {Ritchie},\ and\ \citenamefont {Pepper}}]{AstleyPRL07}%
  \BibitemOpen
  \bibfield  {author} {\bibinfo {author} {\bibfnamefont {M.~R.}\ \bibnamefont
  {Astley}}, \bibinfo {author} {\bibfnamefont {M.}~\bibnamefont {Kataoka}},
  \bibinfo {author} {\bibfnamefont {C.~J.~B.}\ \bibnamefont {Ford}}, \bibinfo
  {author} {\bibfnamefont {C.~H.~W.}\ \bibnamefont {Barnes}}, \bibinfo {author}
  {\bibfnamefont {D.}~\bibnamefont {Anderson}}, \bibinfo {author}
  {\bibfnamefont {G.~A.~C.}\ \bibnamefont {Jones}}, \bibinfo {author}
  {\bibfnamefont {I.}~\bibnamefont {Farrer}}, \bibinfo {author} {\bibfnamefont
  {D.~A.}\ \bibnamefont {Ritchie}}, \ and\ \bibinfo {author} {\bibfnamefont
  {M.}~\bibnamefont {Pepper}},\ }\href {\doibase 10.1103/PhysRevLett.99.156802}
  {\bibfield  {journal} {\bibinfo  {journal} {Phys. Rev. Lett.}\ }\textbf
  {\bibinfo {volume} {99}},\ \bibinfo {pages} {156802} (\bibinfo {year}
  {2007})}\BibitemShut {NoStop}%
\bibitem [{\citenamefont {Kataoka}\ \emph {et~al.}(2009)\citenamefont
  {Kataoka}, \citenamefont {Astley}, \citenamefont {Thorn}, \citenamefont {Oi},
  \citenamefont {Barnes}, \citenamefont {Ford}, \citenamefont {Anderson},
  \citenamefont {Jones}, \citenamefont {Farrer}, \citenamefont {Ritchie},\ and\
  \citenamefont {Pepper}}]{KataokaPRL09}%
  \BibitemOpen
  \bibfield  {author} {\bibinfo {author} {\bibfnamefont {M.}~\bibnamefont
  {Kataoka}}, \bibinfo {author} {\bibfnamefont {M.~R.}\ \bibnamefont {Astley}},
  \bibinfo {author} {\bibfnamefont {A.~L.}\ \bibnamefont {Thorn}}, \bibinfo
  {author} {\bibfnamefont {D.~K.~L.}\ \bibnamefont {Oi}}, \bibinfo {author}
  {\bibfnamefont {C.~H.~W.}\ \bibnamefont {Barnes}}, \bibinfo {author}
  {\bibfnamefont {C.~J.~B.}\ \bibnamefont {Ford}}, \bibinfo {author}
  {\bibfnamefont {D.}~\bibnamefont {Anderson}}, \bibinfo {author}
  {\bibfnamefont {G.~A.~C.}\ \bibnamefont {Jones}}, \bibinfo {author}
  {\bibfnamefont {I.}~\bibnamefont {Farrer}}, \bibinfo {author} {\bibfnamefont
  {D.~A.}\ \bibnamefont {Ritchie}}, \ and\ \bibinfo {author} {\bibfnamefont
  {M.}~\bibnamefont {Pepper}},\ }\href {\doibase
  10.1103/PhysRevLett.102.156801} {\bibfield  {journal} {\bibinfo  {journal}
  {Phys. Rev. Lett.}\ }\textbf {\bibinfo {volume} {102}},\ \bibinfo {pages}
  {156801} (\bibinfo {year} {2009})}\BibitemShut {NoStop}%
\bibitem [{\citenamefont {Buscemi}\ \emph {et~al.}(2009)\citenamefont
  {Buscemi}, \citenamefont {Bordone},\ and\ \citenamefont
  {Bertoni}}]{BuscemiJPCM09}%
  \BibitemOpen
  \bibfield  {author} {\bibinfo {author} {\bibfnamefont {F.}~\bibnamefont
  {Buscemi}}, \bibinfo {author} {\bibfnamefont {P.}~\bibnamefont {Bordone}}, \
  and\ \bibinfo {author} {\bibfnamefont {A.}~\bibnamefont {Bertoni}},\ }\href
  {http://stacks.iop.org/0953-8984/21/i=30/a=305303} {\bibfield  {journal}
  {\bibinfo  {journal} {Journal of Physics: Condensed Matter}\ }\textbf
  {\bibinfo {volume} {21}},\ \bibinfo {pages} {305303} (\bibinfo {year}
  {2009})}\BibitemShut {NoStop}%
\bibitem [{\citenamefont {McNeil}\ \emph {et~al.}(2011)\citenamefont {McNeil},
  \citenamefont {Kataoka}, \citenamefont {Ford}, \citenamefont {Barnes},
  \citenamefont {Anderson}, \citenamefont {Jones}, \citenamefont {Farrer},\
  and\ \citenamefont {Ritchie}}]{McNeil_Nature11}%
  \BibitemOpen
  \bibfield  {author} {\bibinfo {author} {\bibfnamefont {R.~P.~G.}\
  \bibnamefont {McNeil}}, \bibinfo {author} {\bibfnamefont {M.}~\bibnamefont
  {Kataoka}}, \bibinfo {author} {\bibfnamefont {C.~J.~B.}\ \bibnamefont
  {Ford}}, \bibinfo {author} {\bibfnamefont {C.~H.~W.}\ \bibnamefont {Barnes}},
  \bibinfo {author} {\bibfnamefont {D.}~\bibnamefont {Anderson}}, \bibinfo
  {author} {\bibfnamefont {G.~A.~C.}\ \bibnamefont {Jones}}, \bibinfo {author}
  {\bibfnamefont {I.}~\bibnamefont {Farrer}}, \ and\ \bibinfo {author}
  {\bibfnamefont {D.~A.}\ \bibnamefont {Ritchie}},\ }\href {\doibase
  10.1038/nature10444} {\bibfield  {journal} {\bibinfo  {journal} {Nature}\
  }\textbf {\bibinfo {volume} {477}},\ \bibinfo {pages} {439} (\bibinfo {year}
  {2011})}\BibitemShut {NoStop}%
\bibitem [{\citenamefont {Rudolph}\ \emph
  {et~al.}(2007{\natexlab{a}})\citenamefont {Rudolph}, \citenamefont {Hey},\
  and\ \citenamefont {Santos}}]{Rudolph_PRL07}%
  \BibitemOpen
  \bibfield  {author} {\bibinfo {author} {\bibfnamefont {J.}~\bibnamefont
  {Rudolph}}, \bibinfo {author} {\bibfnamefont {R.}~\bibnamefont {Hey}}, \ and\
  \bibinfo {author} {\bibfnamefont {P.~V.}\ \bibnamefont {Santos}},\ }\href
  {\doibase 10.1103/PhysRevLett.99.047602} {\bibfield  {journal} {\bibinfo
  {journal} {Phys. Rev. Lett.}\ }\textbf {\bibinfo {volume} {99}},\ \bibinfo
  {pages} {047602} (\bibinfo {year} {2007}{\natexlab{a}})}\BibitemShut
  {NoStop}%
\bibitem [{\citenamefont {Rudolph}\ \emph
  {et~al.}(2007{\natexlab{b}})\citenamefont {Rudolph}, \citenamefont {Hey},\
  and\ \citenamefont {Santos}}]{Rudolph_ProcCPLMCN07}%
  \BibitemOpen
  \bibfield  {author} {\bibinfo {author} {\bibfnamefont {J.}~\bibnamefont
  {Rudolph}}, \bibinfo {author} {\bibfnamefont {R.}~\bibnamefont {Hey}}, \ and\
  \bibinfo {author} {\bibfnamefont {P.}~\bibnamefont {Santos}},\ }\href
  {\doibase https://doi.org/10.1016/j.spmi.2007.03.008} {\bibfield  {journal}
  {\bibinfo  {journal} {Superlattices and Microstructures}\ }\textbf {\bibinfo
  {volume} {41}},\ \bibinfo {pages} {293 } (\bibinfo {year}
  {2007}{\natexlab{b}})}\BibitemShut {NoStop}%
\bibitem [{\citenamefont {Rosini}\ \emph {et~al.}(2004)\citenamefont {Rosini},
  \citenamefont {Bertoni}, \citenamefont {Bordone},\ and\ \citenamefont
  {Jacoboni}}]{RosiniJCE2004}%
  \BibitemOpen
  \bibfield  {author} {\bibinfo {author} {\bibfnamefont {M.}~\bibnamefont
  {Rosini}}, \bibinfo {author} {\bibfnamefont {A.}~\bibnamefont {Bertoni}},
  \bibinfo {author} {\bibfnamefont {P.}~\bibnamefont {Bordone}}, \ and\
  \bibinfo {author} {\bibfnamefont {C.}~\bibnamefont {Jacoboni}},\ }\href
  {\doibase 10.1007/s10825-004-7093-2} {\bibfield  {journal} {\bibinfo
  {journal} {Journal of Computational Electronics}\ }\textbf {\bibinfo {volume}
  {3}},\ \bibinfo {pages} {443} (\bibinfo {year} {2004})}\BibitemShut {NoStop}%
\bibitem [{\citenamefont {Bordone}\ \emph {et~al.}(2004)\citenamefont
  {Bordone}, \citenamefont {Bertoni}, \citenamefont {Rosini}, \citenamefont
  {Reggiani},\ and\ \citenamefont {Jacoboni}}]{BordoneSST04}%
  \BibitemOpen
  \bibfield  {author} {\bibinfo {author} {\bibfnamefont {P.}~\bibnamefont
  {Bordone}}, \bibinfo {author} {\bibfnamefont {A.}~\bibnamefont {Bertoni}},
  \bibinfo {author} {\bibfnamefont {M.}~\bibnamefont {Rosini}}, \bibinfo
  {author} {\bibfnamefont {S.}~\bibnamefont {Reggiani}}, \ and\ \bibinfo
  {author} {\bibfnamefont {C.}~\bibnamefont {Jacoboni}},\ }\href
  {http://stacks.iop.org/0268-1242/19/i=4/a=135} {\bibfield  {journal}
  {\bibinfo  {journal} {Semiconductor Science and Technology}\ }\textbf
  {\bibinfo {volume} {19}},\ \bibinfo {pages} {S412} (\bibinfo {year}
  {2004})}\BibitemShut {NoStop}%
\bibitem [{\citenamefont {Rodriquez}\ \emph {et~al.}(2005)\citenamefont
  {Rodriquez}, \citenamefont {Oi}, \citenamefont {Kataoka}, \citenamefont
  {Barnes}, \citenamefont {Ohshima},\ and\ \citenamefont
  {Ekert}}]{RodriquezPRB2005}%
  \BibitemOpen
  \bibfield  {author} {\bibinfo {author} {\bibfnamefont {R.}~\bibnamefont
  {Rodriquez}}, \bibinfo {author} {\bibfnamefont {D.~K.~L.}\ \bibnamefont
  {Oi}}, \bibinfo {author} {\bibfnamefont {M.}~\bibnamefont {Kataoka}},
  \bibinfo {author} {\bibfnamefont {C.~H.~W.}\ \bibnamefont {Barnes}}, \bibinfo
  {author} {\bibfnamefont {T.}~\bibnamefont {Ohshima}}, \ and\ \bibinfo
  {author} {\bibfnamefont {A.~K.}\ \bibnamefont {Ekert}},\ }\href {\doibase
  10.1103/PhysRevB.72.085329} {\bibfield  {journal} {\bibinfo  {journal} {Phys.
  Rev. B}\ }\textbf {\bibinfo {volume} {72}},\ \bibinfo {pages} {085329}
  (\bibinfo {year} {2005})}\BibitemShut {NoStop}%
\bibitem [{\citenamefont {Buscemi}\ \emph {et~al.}(2010)\citenamefont
  {Buscemi}, \citenamefont {Bordone},\ and\ \citenamefont
  {Bertoni}}]{BuscemiPRB10}%
  \BibitemOpen
  \bibfield  {author} {\bibinfo {author} {\bibfnamefont {F.}~\bibnamefont
  {Buscemi}}, \bibinfo {author} {\bibfnamefont {P.}~\bibnamefont {Bordone}}, \
  and\ \bibinfo {author} {\bibfnamefont {A.}~\bibnamefont {Bertoni}},\ }\href
  {\doibase 10.1103/PhysRevB.81.045312} {\bibfield  {journal} {\bibinfo
  {journal} {Phys. Rev. B}\ }\textbf {\bibinfo {volume} {81}},\ \bibinfo
  {pages} {045312} (\bibinfo {year} {2010})}\BibitemShut {NoStop}%
\bibitem [{\citenamefont {Shi}\ \emph {et~al.}(2011)\citenamefont {Shi},
  \citenamefont {Zhang},\ and\ \citenamefont {Wei}}]{ShiPRA11}%
  \BibitemOpen
  \bibfield  {author} {\bibinfo {author} {\bibfnamefont {X.}~\bibnamefont
  {Shi}}, \bibinfo {author} {\bibfnamefont {M.}~\bibnamefont {Zhang}}, \ and\
  \bibinfo {author} {\bibfnamefont {L.~F.}\ \bibnamefont {Wei}},\ }\href
  {\doibase 10.1103/PhysRevA.84.062310} {\bibfield  {journal} {\bibinfo
  {journal} {Phys. Rev. A}\ }\textbf {\bibinfo {volume} {84}},\ \bibinfo
  {pages} {062310} (\bibinfo {year} {2011})}\BibitemShut {NoStop}%
\bibitem [{\citenamefont {Grosso}\ \emph {et~al.}(2009)\citenamefont {Grosso},
  \citenamefont {Graves}, \citenamefont {Hammack}, \citenamefont {High},
  \citenamefont {Butov}, \citenamefont {Hanson},\ and\ \citenamefont
  {Gossard}}]{Grosso_NatPhot09}%
  \BibitemOpen
  \bibfield  {author} {\bibinfo {author} {\bibfnamefont {G.}~\bibnamefont
  {Grosso}}, \bibinfo {author} {\bibfnamefont {J.}~\bibnamefont {Graves}},
  \bibinfo {author} {\bibfnamefont {A.~T.}\ \bibnamefont {Hammack}}, \bibinfo
  {author} {\bibfnamefont {A.~A.}\ \bibnamefont {High}}, \bibinfo {author}
  {\bibfnamefont {L.~V.}\ \bibnamefont {Butov}}, \bibinfo {author}
  {\bibfnamefont {M.}~\bibnamefont {Hanson}}, \ and\ \bibinfo {author}
  {\bibfnamefont {A.~C.}\ \bibnamefont {Gossard}},\ }\href {\doibase
  10.1038/nphoton.2009.166} {\bibfield  {journal} {\bibinfo  {journal} {Nature
  Photonics}\ }\textbf {\bibinfo {volume} {3}},\ \bibinfo {pages} {577}
  (\bibinfo {year} {2009})}\BibitemShut {NoStop}%
\bibitem [{\citenamefont {Kinzel}\ \emph {et~al.}(2016)\citenamefont {Kinzel},
  \citenamefont {Schülein}, \citenamefont {Weiß}, \citenamefont {Janker},
  \citenamefont {Bühler}, \citenamefont {Heigl}, \citenamefont {Rudolph},
  \citenamefont {Morkötter}, \citenamefont {Döblinger}, \citenamefont
  {Bichler}, \citenamefont {Abstreiter}, \citenamefont {Finley}, \citenamefont
  {Wixforth}, \citenamefont {Koblmüller},\ and\ \citenamefont
  {Krenner}}]{KinzelACSNano16}%
  \BibitemOpen
  \bibfield  {author} {\bibinfo {author} {\bibfnamefont {J.~B.}\ \bibnamefont
  {Kinzel}}, \bibinfo {author} {\bibfnamefont {F.~J.~R.}\ \bibnamefont
  {Schülein}}, \bibinfo {author} {\bibfnamefont {M.}~\bibnamefont {Weiß}},
  \bibinfo {author} {\bibfnamefont {L.}~\bibnamefont {Janker}}, \bibinfo
  {author} {\bibfnamefont {D.~D.}\ \bibnamefont {Bühler}}, \bibinfo {author}
  {\bibfnamefont {M.}~\bibnamefont {Heigl}}, \bibinfo {author} {\bibfnamefont
  {D.}~\bibnamefont {Rudolph}}, \bibinfo {author} {\bibfnamefont
  {S.}~\bibnamefont {Morkötter}}, \bibinfo {author} {\bibfnamefont
  {M.}~\bibnamefont {Döblinger}}, \bibinfo {author} {\bibfnamefont
  {M.}~\bibnamefont {Bichler}}, \bibinfo {author} {\bibfnamefont
  {G.}~\bibnamefont {Abstreiter}}, \bibinfo {author} {\bibfnamefont {J.~J.}\
  \bibnamefont {Finley}}, \bibinfo {author} {\bibfnamefont {A.}~\bibnamefont
  {Wixforth}}, \bibinfo {author} {\bibfnamefont {G.}~\bibnamefont
  {Koblmüller}}, \ and\ \bibinfo {author} {\bibfnamefont {H.~J.}\ \bibnamefont
  {Krenner}},\ }\href {\doibase 10.1021/acsnano.5b07639} {\bibfield  {journal}
  {\bibinfo  {journal} {ACS Nano}\ }\textbf {\bibinfo {volume} {10}},\ \bibinfo
  {pages} {4942} (\bibinfo {year} {2016})},\ \bibinfo {note} {pMID: 27007813},\
  \Eprint {http://arxiv.org/abs/http://dx.doi.org/10.1021/acsnano.5b07639}
  {http://dx.doi.org/10.1021/acsnano.5b07639} \BibitemShut {NoStop}%
\bibitem [{\citenamefont {Grasselli}\ \emph {et~al.}(2015)\citenamefont
  {Grasselli}, \citenamefont {Bertoni},\ and\ \citenamefont
  {Goldoni}}]{GrasselliJCP15}%
  \BibitemOpen
  \bibfield  {author} {\bibinfo {author} {\bibfnamefont {F.}~\bibnamefont
  {Grasselli}}, \bibinfo {author} {\bibfnamefont {A.}~\bibnamefont {Bertoni}},
  \ and\ \bibinfo {author} {\bibfnamefont {G.}~\bibnamefont {Goldoni}},\ }\href
  {\doibase http://dx.doi.org/10.1063/1.4905483} {\bibfield  {journal}
  {\bibinfo  {journal} {The Journal of Chemical Physics}\ }\textbf {\bibinfo
  {volume} {142}},\ \bibinfo {pages} {034701} (\bibinfo {year}
  {2015})}\BibitemShut {NoStop}%
\bibitem [{\citenamefont {Grasselli}\ \emph
  {et~al.}(2016{\natexlab{a}})\citenamefont {Grasselli}, \citenamefont
  {Bertoni},\ and\ \citenamefont {Goldoni}}]{GrasselliPRB16}%
  \BibitemOpen
  \bibfield  {author} {\bibinfo {author} {\bibfnamefont {F.}~\bibnamefont
  {Grasselli}}, \bibinfo {author} {\bibfnamefont {A.}~\bibnamefont {Bertoni}},
  \ and\ \bibinfo {author} {\bibfnamefont {G.}~\bibnamefont {Goldoni}},\ }\href
  {\doibase {10.1103/PhysRevB.93.195310}} {\bibfield  {journal} {\bibinfo
  {journal} {Physical Review B}\ }\textbf {\bibinfo {volume} {93}},\ \bibinfo
  {pages} {195310} (\bibinfo {year} {2016}{\natexlab{a}})}\BibitemShut
  {NoStop}%
\bibitem [{\citenamefont {Grasselli}\ \emph
  {et~al.}(2016{\natexlab{b}})\citenamefont {Grasselli}, \citenamefont
  {Bertoni},\ and\ \citenamefont {Goldoni}}]{GrasselliJPCS16}%
  \BibitemOpen
  \bibfield  {author} {\bibinfo {author} {\bibfnamefont {F.}~\bibnamefont
  {Grasselli}}, \bibinfo {author} {\bibfnamefont {A.}~\bibnamefont {Bertoni}},
  \ and\ \bibinfo {author} {\bibfnamefont {G.}~\bibnamefont {Goldoni}},\ }in\
  \href@noop {} {\emph {\bibinfo {booktitle} {Journal of Physics: Conference
  Series}}},\ Vol.\ \bibinfo {volume} {738}\ (\bibinfo {organization} {IOP
  Publishing},\ \bibinfo {year} {2016})\ p.\ \bibinfo {pages}
  {012028}\BibitemShut {NoStop}%
\bibitem [{\citenamefont {Grasselli}\ \emph
  {et~al.}(2016{\natexlab{c}})\citenamefont {Grasselli}, \citenamefont
  {Bertoni},\ and\ \citenamefont {Goldoni}}]{GrasselliPRB16b}%
  \BibitemOpen
  \bibfield  {author} {\bibinfo {author} {\bibfnamefont {F.}~\bibnamefont
  {Grasselli}}, \bibinfo {author} {\bibfnamefont {A.}~\bibnamefont {Bertoni}},
  \ and\ \bibinfo {author} {\bibfnamefont {G.}~\bibnamefont {Goldoni}},\ }\href
  {\doibase 10.1103/PhysRevB.94.125418} {\bibfield  {journal} {\bibinfo
  {journal} {Phys. Rev. B}\ }\textbf {\bibinfo {volume} {94}},\ \bibinfo
  {pages} {125418} (\bibinfo {year} {2016}{\natexlab{c}})}\BibitemShut
  {NoStop}%
\bibitem [{\citenamefont {{Federico Grasselli and Andrea Bertoni and Guido
  Goldoni}}(2017)}]{GrasselliSLMS17}%
  \BibitemOpen
  \bibfield  {author} {\bibinfo {author} {\bibnamefont {{Federico Grasselli and
  Andrea Bertoni and Guido Goldoni}}},\ }\href {\doibase
  {https://doi.org/10.1016/j.spmi.2016.12.018}} {\bibfield  {journal} {\bibinfo
   {journal} {{Superlattices and Microstructures }}\ }\textbf {\bibinfo
  {volume} {108}},\ \bibinfo {pages} {73} (\bibinfo {year}
  {{2017}})}\BibitemShut {NoStop}%
\bibitem [{\citenamefont {Lobanov}\ \emph {et~al.}(2016)\citenamefont
  {Lobanov}, \citenamefont {Gippius},\ and\ \citenamefont
  {Butov}}]{Lobanov_PRB16}%
  \BibitemOpen
  \bibfield  {author} {\bibinfo {author} {\bibfnamefont {S.~V.}\ \bibnamefont
  {Lobanov}}, \bibinfo {author} {\bibfnamefont {N.~A.}\ \bibnamefont
  {Gippius}}, \ and\ \bibinfo {author} {\bibfnamefont {L.~V.}\ \bibnamefont
  {Butov}},\ }\href {\doibase 10.1103/PhysRevB.94.245401} {\bibfield  {journal}
  {\bibinfo  {journal} {Phys. Rev. B}\ }\textbf {\bibinfo {volume} {94}},\
  \bibinfo {pages} {245401} (\bibinfo {year} {2016})}\BibitemShut {NoStop}%
\bibitem [{\citenamefont {{R. Zimmermann and F. Gro{\ss}e and E.
  Runge}}(1997)}]{ZimmermannPAC97}%
  \BibitemOpen
  \bibfield  {author} {\bibinfo {author} {\bibnamefont {{R. Zimmermann and F.
  Gro{\ss}e and E. Runge}}},\ }\href {\doibase
  {http://dx.doi.org/10.1351/pac199769061179}} {\bibfield  {journal} {\bibinfo
  {journal} {{Pure Appl. Chem.}}\ }\textbf {\bibinfo {volume} {69}},\ \bibinfo
  {pages} {1179 } (\bibinfo {year} {1997})}\BibitemShut {NoStop}%
\bibitem [{\citenamefont {Combescot}\ \emph {et~al.}(2017)\citenamefont
  {Combescot}, \citenamefont {Combescot},\ and\ \citenamefont
  {Dubin}}]{Combescot_RepProgPhys17}%
  \BibitemOpen
  \bibfield  {author} {\bibinfo {author} {\bibfnamefont {M.}~\bibnamefont
  {Combescot}}, \bibinfo {author} {\bibfnamefont {R.}~\bibnamefont
  {Combescot}}, \ and\ \bibinfo {author} {\bibfnamefont {F.}~\bibnamefont
  {Dubin}},\ }\href {http://stacks.iop.org/0034-4885/80/i=6/a=066501}
  {\bibfield  {journal} {\bibinfo  {journal} {Reports on Progress in Physics}\
  }\textbf {\bibinfo {volume} {80}},\ \bibinfo {pages} {066501} (\bibinfo
  {year} {2017})}\BibitemShut {NoStop}%
\bibitem [{Note1()}]{Note1}%
  \BibitemOpen
  \bibinfo {note} {{The quantity {{$\delimiter "426830A X \delimiter "526930B
  (t)$}} is the expectation value for quantum simulations, while it is the
  average over the different trajectories for the classical evolution, as a
  function of time $t$.}}\BibitemShut {Stop}%
\bibitem [{Note2()}]{Note2}%
  \BibitemOpen
  \bibinfo {note} {To avoid non-physical values we only accumulate the averages
  when $\rho _\protect \mathrm {c.m.}(X;t) > \geq 5$\% of its total value. This
  explains, for instance, why the dashed line starts at an intermediate time
  and not from $t=0$.}\BibitemShut {Stop}%
\bibitem [{Note3()}]{Note3}%
  \BibitemOpen
  \bibinfo {note} {In Fig.~\ref {fig:SAW:figure_diff_ampl_orizz}(b) we actually
  reported Fig.~\ref {fig:SAW:figure2}(b), even if the vertical axis is
  rescaled, to allow the full representation of the potential in Fig.~\ref
  {fig:SAW:figure_diff_ampl_orizz}(c), keeping the same scale in the three
  panels.}\BibitemShut {Stop}%
\bibitem [{\citenamefont {Butov}(2017)}]{Butov_SLAMS17}%
  \BibitemOpen
  \bibfield  {author} {\bibinfo {author} {\bibfnamefont {L.}~\bibnamefont
  {Butov}},\ }\href {\doibase https://doi.org/10.1016/j.spmi.2016.12.035}
  {\bibfield  {journal} {\bibinfo  {journal} {Superlattices and
  Microstructures}\ }\textbf {\bibinfo {volume} {108}},\ \bibinfo {pages} {2 }
  (\bibinfo {year} {2017})},\ \bibinfo {note} {indirect Excitons: Physics and
  Applications}\BibitemShut {NoStop}%
\end{thebibliography}

%

\end{document}